\documentclass[12pt]{article}
\usepackage{amsmath,amssymb,amsthm,amsxtra,bbm,overpic,bm,epsfig,subfigure}
\usepackage[colorlinks]{hyperref}
\usepackage{mathrsfs}
\usepackage{enumitem}
\usepackage{graphicx}
\usepackage{color}
\usepackage{comment}
\usepackage{epstopdf}
\usepackage{float}
\usepackage{cite}
\usepackage{upgreek,overpic,textcomp,multirow}
\numberwithin{equation}{section}
\allowdisplaybreaks[3]

\usepackage{hyperref}
\usepackage{slashed,stmaryrd}

\def\thefootnote{\fnsymbol{footnote}}

\begin{document}
	
	\vspace{0.2cm}
	
	\begin{center}
		{\Large\bf Helicity-changing Decays of Cosmological Relic Neutrinos}
	\end{center}
	
	\vspace{0.2cm}
	
	\begin{center}
		{\bf Jihong Huang}~\footnote{E-mail: huangjh@ihep.ac.cn},
		\quad
		{\bf Shun Zhou}~\footnote{E-mail: zhoush@ihep.ac.cn (corresponding author)}
		\\
		\vspace{0.2cm}
		{\small
			Institute of High Energy Physics, Chinese Academy of Sciences, Beijing 100049, China\\
			School of Physical Sciences, University of Chinese Academy of Sciences, Beijing 100049, China}
	\end{center}
	
	\vspace{1.5cm}
	
\begin{abstract}
In this paper, we examine the possibility that massive neutrinos are unstable due to their invisible decays $\nu^{}_i \to \nu^{}_j + \phi$, where $\nu^{}_i$ and $\nu^{}_j$ (for $i, j = 1, 2, 3$) are any two of neutrino mass eigenstates with masses $m^{}_i > m^{}_j$ and $\phi$ is a massless Nambu-Goldstone boson, and explore the implications for the detection of cosmological relic neutrinos in the present Universe. First, we carry out a complete calculation of neutrino decay rates in the general case where the individual helicities of parent and daughter neutrinos are specified. Then, the invisible decays of cosmological relic neutrinos are studied and their impact on the capture rates on the beta-decaying nuclei (e.g., $\nu^{}_e + {^3{\rm H}} \to {^3{\rm He}} + e^-$) is analyzed. The invisible decays of massive neutrinos could substantially change the capture rates in the PTOLEMY-like experiments when compared to the case of stable neutrinos. In particular, we find that the helicity-changing decays of Dirac neutrinos play an important role whereas those of Majorana neutrinos have no practical effects. However, if a substantial fraction of heavier neutrinos decay into the lightest one, the detection of relic neutrinos will require a much higher energy resolution and thus be even more challenging.
\end{abstract}
	
	\newpage
	
	\def\thefootnote{\arabic{footnote}}
	\setcounter{footnote}{0}
	
\section{Introduction}\label{sec:intro}
	
Though neutrino oscillation experiments have provided us with robust evidence that neutrinos are massive, the intrinsic properties of massive neutrinos are largely unknown~\cite{ParticleDataGroup:2022pth,Xing:2020ijf}. It remains to be determined whether three neutrino masses $m^{}_i$ (for $i = 1, 2, 3$) take on the normal ordering (NO), i.e., $m^{}_1 < m^{}_2 < m^{}_3$, or the inverted ordering (IO), i.e., $m^{}_3 < m^{}_1 < m^{}_2$. In addition, the absolute scale of neutrino masses is not yet pinned down such that the lightest neutrino can in principle be massless. More importantly, massive neutrinos can be either Dirac or Majorana particles, and the lepton number conservation must be violated in the latter case. Now that at least two species of neutrinos are massive, it is also interesting to investigate whether a heavier neutrino can decay into a lighter one and other elementary particles within or beyond the Standard Model (SM).

In the minimal extension of the SM with a nonzero neutrino mass term, one can immediately realize that radiative neutrino decays $\nu^{}_i \to \nu^{}_j + \gamma$ with $m^{}_i > m^{}_j$ indeed take place at the one-loop level. However, the rates of such radiative decays are highly suppressed due to the small effective magnetic moment of massive neutrinos~\cite{Fujikawa:1980yx} so that the lifetimes of ordinary neutrinos are much longer than the age of the Universe. In a class of neutrino mass models~\cite{Gelmini:1980re,Barger:1981vd,Valle:1983ua,Gonzalez-Garcia:1988okv}, neutrinos turn out to interact with the Nambu-Goldstone boson~\cite{Nambu:1960xd, Goldstone:1961eq}, i.e., the Majoron, arising from the spontaneous breakdown of the lepton number conservation. In order to make a meaningful comparison between Dirac and Majorana neutrino decays, we adopt the following phenomenological Lagrangian for Dirac neutrinos
\begin{eqnarray} \label{eq:LagD}
	{\cal L}^{}_{\rm D} = \sum_i \left(\overline{\nu^{}_i} {\rm i}\slashed{\partial} \nu^{}_i - m^{}_i \overline{\nu^{}_i} \nu^{}_i\right) + \frac{1}{2} \partial^{}_\mu \phi \partial^\mu \phi - \left[ {\rm i}\phi \sum_{i, j} g^{}_{ij} \overline{\nu^{}_i} \gamma^5 \nu^{}_j + {\rm h.c.} \right] \; ;
\end{eqnarray}
and that for Majorana neutrinos
\begin{eqnarray} \label{eq:LagM}
	{\cal L}^{}_{\rm M} = \frac{1}{2} \sum_i \left(\overline{\nu^{}_i} {\rm i}\slashed{\partial} \nu^{}_i - m^{}_i \overline{\nu^{}_i} \nu^{}_i \right) + \frac{1}{2} \partial^{}_\mu \phi \partial^\mu \phi - \left[ {\rm i}\phi \sum_{i, j} g^{}_{ij} \overline{\nu^{}_i} \gamma^5 \nu^{}_j + {\rm h.c.} \right] \; ,
\end{eqnarray}
where $\phi$ stands for a massless pseudo-scalar boson and the coupling constants $g^{}_{ij}$ (for $i, j = 1, 2, 3$) are assumed to be real for simplicity. In the case of Dirac neutrinos, the interaction term and its Hermitian conjugate in the square brackets in Eq.~(\ref{eq:LagD}) induce the decays of antineutrinos $\overline{\nu}^{}_i \to \overline{\nu}^{}_j + \phi$ and those of neutrinos $\nu^{}_i \to \nu^{}_j + \phi$, respectively. In the case of Majorana neutrinos, the interaction term and its Hermitian conjugate in the square brackets in Eq.~(\ref{eq:LagM}) are identical, leading to the decay amplitude of $\nu^{}_i \to \nu^{}_j + \phi$ for Majorana neutrinos twice that for Dirac ones given the same coupling constants. 

In this paper, we examine these invisible decays of massive Dirac or Majorana neutrinos, and explore the impact of helicity-changing decays on the capture rates of cosmological relic neutrinos on the beta-decaying nuclei. The motivation of such a study is two-fold. First, being a prominent prediction of the standard cosmology, the relic neutrinos from the Big Bang definitely exist in the present Universe as a cosmic neutrino background (C$\nu$B). The experimental detection of the C$\nu$B will not only offer unambiguous evidence for the correctness of the Big Bang theory, but also novel opportunities to probe the intrinsic properties of massive neutrinos, such as their mass ordering~\cite{Blennow:2008fh}, flavor mixing~\cite{Li:2011ne}, Majorana nature~\cite{Long:2014zva, Chen:2015dka, Zhang:2015wua, Huang:2016qmh, Roulet:2018fyh} and lifetimes~\cite{Akita:2021hqn}. Second, if neutrinos decay gradually, their number density in the present Universe and the capture rate of C$\nu$B on the beta-decaying nuclei~\cite{Weinberg:1962zza}, such as $\nu_e^{} + {}^3{\rm H} \to {}^3{\rm He} + e^-$, will be quite different. For instance, the Princeton Tritium Observatory for Light, Early-Universe, Massive-Neutrino Yield (PTOLEMY) experiment aims to detect C$\nu$B with $100~{\rm g}$ tritium source deposited onto the graphene substrate~\cite{Betts:2013uya}, where the capture rate for Dirac neutrinos is $\Gamma_{\rm C\nu B}^{\rm D} \approx 4~{\rm yr}^{-1}$, while for Majorana neutrinos it is twice larger, i.e., $\Gamma_{\rm C\nu B}^{\rm M} \approx 8~{\rm yr}^{-1}$~\cite{Long:2014zva}. However, after taking account of the momentum distribution of relic neutrinos in the Dirac case, the authors of Ref.~\cite{Roulet:2018fyh} have found that the ratio $\Gamma_{\rm C\nu B}^{\rm D}/\Gamma_{\rm C\nu B}^{\rm M}$ increases from 0.5 to 0.84 when the lightest neutrino is nearly massless in the NO case. Therefore, it is intriguing to see how the invisible decays of massive neutrinos modify the capture rates.
	
The remaining part of this paper is organized as follows. In Sec.~\ref{sec:decay}, we calculate the decay rates of massive neutrinos with specific helicities in the laboratory frame. Current observational constraints on neutrino lifetimes are summarized as well. Then, the evolution of decaying neutrinos in the Universe is discussed and the capture rates of C$\nu$B are computed in Sec.~\ref{sec:cnb}, where both scenarios of Majorana and Dirac neutrinos in the cases of different mass orderings are considered. Finally, we summarize our main results and conclude in Sec.~\ref{sec:sum}.

\section{Invisible Decays of Massive Neutrinos}\label{sec:decay}
	
\subsection{Helicity-changing Decays}
	
A detailed calculation of the decay amplitudes for massive neutrinos interacting with either scalar or pseudo-scalar bosons can be found in Refs.~\cite{Kim:1990km, Funcke:2019grs}. In the subsequent discussions in this section, we closely follow the notations in the latter of these two references.
	
Given the Lagrangian for massive neutrinos in Eq.~(\ref{eq:LagD}) or Eq.~(\ref{eq:LagM}), one can immediately write down the decay amplitudes for $\nu^{}_i(p^{}_i, h^{}_i) \to \nu^{}_j(p^{}_j, h^{}_j) + \phi(k)$, where $p^{}_i = (E^{}_i, {\bf p}^{}_i)$, $p^{}_j = (E^{}_j, {\bf p}^{}_j)$ and $k = (\omega, {\bf k})$ refer to the four-momenta of parent and daughter particles and $h^{}_i$ (or $h^{}_j$) denotes the helicity of $\nu^{}_i$ (or $\nu^{}_j$). More explicitly, the decay amplitudes for Majorana neutrinos read
\begin{eqnarray} \label{eq:amplitude}
	{\rm i}{\cal M}^{\rm M}_{h_i^{}h_j^{},ij} = 2g_{ij}^{} \overline{u_j^{}}(p_j^{},h_j^{}) \gamma_{}^5 u_i^{}(p_i^{},h_i^{}) \;,
\end{eqnarray}
and we have ${\cal M}^{\rm D}_{h_i^{}h_j^{},ij} = {\cal M}^{\rm M}_{h_i^{}h_j^{},ij}/2$ for Dirac neutrinos. In the Majorana case, neutrinos and antineutrinos are actually indistinguishable in the sense of $\nu^{}_i = \nu^{\rm C}_i$, where the charge-conjugate field $\nu^{\rm C}_i \equiv {\cal C}\overline{\nu^{}_i}^{\rm T}$ is defined with ${\cal C} \equiv {\rm i}\gamma^2\gamma^0$ being the charge-conjugation matrix.

\subsubsection{Decay Amplitudes}

We first recall the standard wave functions of free Dirac fermions. Taking the spin quantization to be along the positive direction of $z$-axis, we have the helicity operator defined as $\widehat{h} \equiv \widehat{\bf p}\cdot\boldsymbol{\sigma}$ with $\widehat{\bf p}\equiv {\bf p}/|{\bf p}| = (\sin\theta\cos\phi,\sin\theta\sin\phi,\cos\theta)$ being the direction of three-momentum ${\bf p}$. In this case, the helicity spinors with $h = +1$ and $-1$ are given by~\cite{Dreiner:2008tw}
\begin{eqnarray}
	\chi_{}^{(+)}(\theta,\phi) = \begin{pmatrix} \displaystyle \cos\frac{\theta}{2} \\ \displaystyle {\rm e}^{+{\rm i} \phi}\sin\frac{\theta}{2} \end{pmatrix} \;, \quad \chi_{}^{(-)}(\theta,\phi) = \begin{pmatrix} \displaystyle -\sin\frac{\theta}{2} \\ \displaystyle {\rm e}^{+{\rm i} \phi}\cos\frac{\theta}{2} \end{pmatrix} \;,
\end{eqnarray}
which satisfy the normalization condition $\chi^{(h)\dag}_{} \chi^{(h')}_{} = \delta^{hh'}$. With these spinors and the conventions adopted in Ref.~\cite{Peskin:1995ev}, the wave functions can be expressed as
\begin{eqnarray} \label{eq:explicit_spinor_+}
	u(p, h = +1) = \cos\frac{\theta}{2} \begin{pmatrix}
		\sqrt{E-|{\bf p}|}  \\ 0 \\
		\sqrt{E+|{\bf p}|}  \\ 0  \end{pmatrix} + \sin\frac{\theta}{2} {\rm e}^{{\rm i}\phi} \begin{pmatrix}
		0 \\ \sqrt{E+|{\bf p}|}  \\
		0 \\ \sqrt{E-|{\bf p}|} \end{pmatrix}  \;,
\end{eqnarray}
for the positive helicity, and 
\begin{eqnarray} \label{eq:explicit_spinor_-}
	u(p, h = -1) = - \sin\frac{\theta}{2} \begin{pmatrix}
		\sqrt{E+|{\bf p}|} \\ 0 \\
		\sqrt{E-|{\bf p}|} \\ 0 \end{pmatrix} + \cos\frac{\theta}{2} {\rm e}^{{\rm i}\phi} \begin{pmatrix}
		0\\ \sqrt{E-|{\bf p}|}  \\
		0\\ \sqrt{E+|{\bf p}|} \end{pmatrix}  \;,
\end{eqnarray}
for the negative helicity. It is worth stressing that both positive- and negative-helical states of daughter neutrinos can be produced in the decays, regardless of the direction of the three-momentum ${\bf p}_j^{}$. For example, if we set ${\bf p}_i^{}$ of the parent neutrino to be along the positive direction of $z$-axis, which is also the spin direction, then its wave function with $h_i^{} = +1$ can be immediately read off from Eq.~(\ref{eq:explicit_spinor_+}), i.e.,
\begin{eqnarray} \label{eq:u_i_+}
	u^{}_{i} (p^{}_i, h^{}_i = +1) = \begin{pmatrix}
		\sqrt{E^{}_i -|{\bf p}^{}_i|}  \\ 0 \\ \sqrt{E^{}_i +|{\bf p}^{}_i|}  \\ 0 \end{pmatrix} \;.
\end{eqnarray}
Based on the orthogonality condition for the wave functions, one can observe that the terms proportional to $\cos(\theta_j^{}/2)$ in Eq.~(\ref{eq:explicit_spinor_+}) and $\sin(\theta^{}_j/2)$ in Eq.~(\ref{eq:explicit_spinor_-}), when applied to the final-state neutrino $\nu^{}_j(p^{}_j, h^{}_j)$, contribute to the decay amplitudes for $h_j^{} = + 1$ and $h^{}_j = -1$, respectively. 

Then, we proceed with the amplitude squared, which can be easily found from Eq.~(\ref{eq:amplitude})
\begin{eqnarray} \label{eq:M^2_tr}
	|{\cal M}_{h_i^{}h_j^{},ij}^{\rm M}|^2 = 4g_{ij}^2 {\rm Tr}\left[u_i^{}(p^{}_i, h^{}_i) \overline{u_i^{}(p^{}_i, h^{}_i)} \gamma^5_{} u_j^{}(p^{}_j, h^{}_j) \overline{u_j^{}(p^{}_j, h^{}_j)} \gamma^5_{} \right] \;,
\end{eqnarray}
for Majorana neutrinos, while $|{\cal M}_{h_i^{}h_j^{},ij}^{\rm D}|^2 = |{\cal M}_{h_i^{}h_j^{},ij}^{\rm M}|^2 / 4$ for Dirac neutrinos. Introducing the polarization four-vector $s^\mu = (|{\bf p}|/m, E\widehat{\bf p}/m)$ for a massive fermion, one can in general obtain
\begin{eqnarray}
	u(p, h)\overline{u(p, h)} = \frac{1}{2}(\slashed{p} + m)(1 + h \gamma_{}^5 \slashed{s})\; ,
\end{eqnarray}
which together with Eq.~(\ref{eq:M^2_tr}) leads to
\begin{eqnarray}
	|{\cal M}_{h_i^{}h_j^{},ij}^{\rm M}|^2 = 4 g_{ij}^2 \left[(1+ h_i^{} h_j^{} s_i^{}\cdot s_j^{}) (p_i^{}\cdot p_j^{}-m_i^{} m_j^{})-h_i^{} h_j^{} (p_i^{}\cdot s_j^{}) (p_j^{}\cdot s_i^{}) \right] \;.
\end{eqnarray} 
After specifying all the four-momenta, masses and polarization vectors of parent and daughter neutrinos, we have
\begin{eqnarray} \label{eq:M^2_E_p}
	|{\cal M}_{h_i^{}h_j^{},ij}^{\rm M}|^2 = 4g_{ij}^2  (E_i^{} E_j^{}-m_i^{} m_j^{}-h_i^{} h_j^{} |{\bf p}_i^{}| |{\bf p}_j^{}|) (1+ h_i^{} h_j^{} \cos\theta_j^{} ) \;.
\end{eqnarray}
It is obvious from Eq.~(\ref{eq:M^2_E_p}) that the squared amplitude for helicity-preserving decay with $h_i^{} h_j^{} = +1$ is proportional to $\cos^2(\theta_j^{}/2)$, while that for helicity-changing decay with $h^{}_i h^{}_j = -1$ to $\sin^2(\theta_j^{}/2)$. This observation is in accordance with the previous analysis below Eq.~(\ref{eq:u_i_+}).

On the other hand, with the explicit expressions of the wave functions in Eqs.~(\ref{eq:explicit_spinor_+})-(\ref{eq:u_i_+}), we can also substitute them into Eq.~(\ref{eq:M^2_tr}) and derive the squared amplitudes 
\begin{eqnarray} 
	\left|{\cal M}_{++,ij}^{\rm M}\right|^2 &=& 8g_{ij}^2 (E_i^{} E_j^{}-m_i^{} m_j^{}- |{\bf p}_i^{}| |{\bf p}_j^{}|) \cos^2\frac{\theta_j^{}}{2} \;, \label{eq:M^2_++}\\
	\left|{\cal M}_{+-,ij}^{\rm M}\right|^2 &=& 8g_{ij}^2 (E_i^{} E_j^{}-m_i^{} m_j^{} + |{\bf p}_i^{}| |{\bf p}_j^{}|) \sin^2\frac{\theta_j^{}}{2} \;, \label{eq:M^2_+-}
\end{eqnarray}
which are the same as those in Eq.~(\ref{eq:M^2_E_p}) for $h_i^{} h_j^{} = +1$ and $h_i^{} h_j^{} = -1$, respectively. For $h_i^{}  =-1$, where the parent neutrino is traveling along the negative direction of $z$-axis, the wave function turns out to be 
\begin{eqnarray} \label{eq:u_i_-}
	u^{}_{i} (p^{}_i, h^{}_i = -1) = - \begin{pmatrix}
		\sqrt{E^{}_i +|{\bf p}^{}_i|} \\ 0 \\
		\sqrt{E^{}_i -|{\bf p}^{}_i|} \\ 0 \end{pmatrix}\;.
\end{eqnarray}
Since the momentum direction of ${\bf p}_i^{}$ is reversed in this case (namely, the polar angle $\theta^{}_i = \pi$), one has to replace $\theta^{}_j$ by $\pi - \theta^{}_j$ in the wave functions of daughter neutrinos in the decay amplitudes. As a consequence, the decay amplitude for $h^{}_i = -1$ and $h_j^{}  =+1$ becomes proportional to $\cos[(\pi-\theta_j^{})/2] = \sin(\theta^{}_j/2)$,  whereas that for $h^{}_i = -1$ and $h_j^{}  =+1$ to $\sin[(\pi-\theta_j^{})/2] = \cos(\theta_j^{}/2)$. Such a correspondence indicates the identities $|{\cal M}_{++,ij}^{\rm M}|^2 = |{\cal M}_{--,ij}^{\rm M}|^2$ and $|{\cal M}_{+-,ij}^{\rm M}|^2=|{\cal M}_{-+,ij}^{\rm M}|^2$ for the squared amplitudes. Certainly these identities can also be obtained from Eq.~(\ref{eq:M^2_E_p}).

\subsubsection{Decay Rates}

After integrating over the phase space of final-state particles, the differential rates for the helicity-preserving decays with $h^{}_i = h^{}_j = \pm 1$ in the Majorana case can be written as
\begin{eqnarray} \label{eq:dGammaMajp}
	\frac{{\rm d}\Gamma^{\rm M}_{\pm\pm,ij}}{{\rm d}E^{}_j} =  \frac{g^2_{ij}m^2_i}{4\pi E^{}_i} \left[+\frac{E^2_i r^2_{ji} + E^2_j}{|{\bf p}^{}_i|^2 |{\bf p}^{}_j|} - \frac{(1 + r^{}_{ji})^2}{2|{\bf p}^{}_i|} \frac{E^{}_i E^{}_j}{|{\bf p}^{}_i| |{\bf p}^{}_j|} + \frac{(1 - r^{}_{ji})^2}{2|{\bf p}^{}_i| } \left(1 + \frac{m^2_i r^{}_{ji}}{|{\bf p}^{}_i| |{\bf p}^{}_j|}\right)\right] \; ,
\end{eqnarray}
with $0 \leqslant r^{}_{ji} \equiv m^{}_j /m^{}_i < 1$ being the neutrino mass ratio; while those for the helicity-changing decays $\nu^{}_i(p^{}_i, h^{}_i) \to \nu^{}_j(p^{}_j, h^{}_j) + \phi(k)$ with $h^{}_i = \pm 1$ and $h^{}_j = \mp 1$ are given by
\begin{eqnarray} \label{eq:dGammaMajm}
	\frac{{\rm d}\Gamma^{\rm M}_{\pm\mp,ij}}{{\rm d}E^{}_j} =  \frac{g^2_{ij}m^2_i}{4\pi E^{}_i} \left[-\frac{E^2_i r^2_{ji} + E^2_j}{|{\bf p}^{}_i|^2 |{\bf p}^{}_j|} + \frac{(1 + r^{}_{ji})^2}{2|{\bf p}^{}_i|} \frac{E^{}_i E^{}_j}{|{\bf p}^{}_i| |{\bf p}^{}_j|} + \frac{(1 - r^{}_{ji})^2}{2|{\bf p}^{}_i| } \left(1 - \frac{m^2_i r^{}_{ji}}{|{\bf p}^{}_i| |{\bf p}^{}_j|}\right)\right] \; .
\end{eqnarray}
In the case of Dirac neutrinos, one can derive the differential decay rates via the following relations
\begin{eqnarray}
	\frac{{\rm d}\Gamma^{\rm D}_{\pm\pm,ij}}{{\rm d}E^{}_j} = \frac{{\rm d}\overline{\Gamma}^{\rm D}_{\mp\mp,ij}}{{\rm d}E^{}_j} = \frac{1}{4} \frac{{\rm d}\Gamma^{\rm M}_{\pm\pm,ij}}{{\rm d}E^{}_j} \; , \quad \frac{{\rm d}\Gamma^{\rm D}_{\pm\mp,ij}}{{\rm d}E^{}_j} = \frac{{\rm d}\overline{\Gamma}^{\rm D}_{\mp\pm,ij}}{{\rm d}E^{}_j} = \frac{1}{4} \frac{{\rm d}\Gamma^{\rm M}_{\pm\mp,ij}}{{\rm d}E^{}_j} \; ,
	\label{eq:dGammaDir}
\end{eqnarray}
where $\Gamma^{\rm D}_{\pm\pm,ij}$ and $\overline{\Gamma}^{\rm D}_{\pm\pm,ij}$ refer to the decay rates for neutrinos $\nu^{}_i(p^{}_i, h^{}_i) \to \nu^{}_j(p^{}_j, h^{}_j) + \phi(k)$ and those for antineutrinos $\overline{\nu}^{}_i(p^{}_i, h^{}_i) \to \overline{\nu}^{}_j(p^{}_j, h^{}_j) + \phi(k)$ with $h^{}_i = h^{}_j = \pm 1$, respectively, and likewise for the helicity-changing decay rates $\Gamma^{\rm D}_{\pm\mp,ij}$ and $\overline{\Gamma}^{\rm D}_{\pm\mp,ij}$. Note that the relations between the decay rates of neutrinos and those of antineutrinos in the case of Dirac neutrinos can be verified by requiring either CP or CPT invariance.
	
	By integrating the differential rates in Eqs.~(\ref{eq:dGammaMajp})-(\ref{eq:dGammaDir}) over the final-state neutrino energy $E^{}_j$, we can derive the most general formulae for the total decay rates for a given configuration of neutrino helicities. Before doing so, one should figure out the maximal and minimal values of $E^{}_j$ from the energy-momentum conservation for two-body decays in the laboratory frame (LF). More explicitly, we have
	\begin{eqnarray}
		E^{\rm max}_j &=& \frac{E^{}_i}{2} \left(1+r^2_{ji}\right) + \frac{|{\bf p}^{}_i|}{2} \left(1 - r^2_{ji}\right) \; , \label{eq:Ejmax} \\
		E^{\rm min}_j &=& \frac{E^{}_i}{2} \left(1+r^2_{ji}\right) - \frac{|{\bf p}^{}_i|}{2} \left(1 - r^2_{ji}\right) \; . \label{eq:Ejmin} 
	\end{eqnarray}
	Then it is straightforward to obtain the total helicity-preserving and -changing decay rates in the LF for a specific four-momentum of the parent neutrino $(E^{}_i, {\bf p}^{}_i)$, namely,
	\begin{eqnarray}
		\Gamma^{\rm M}_{\pm\pm,ij} &=& \frac{g^2_{ij}m^{}_i}{4\pi} \sqrt{1 - \beta^2_i} \left\{ \frac{1}{2} \left(1 - r^2_{ji}\right) (1 - r^{}_{ji})^2 - r^{}_{ji} (1 - r^2_{ji}) - 2r^2_{ji} \ln r^{}_{ji} \right. \nonumber \\
		&~& \left. - (\beta^{-2}_i - 1) \left[ (1+r^{}_{ji}+r^2_{ji}) r^{}_{ji} \ln r^{}_{ji} + \frac{1}{4} (1 - r^2_{ji}) (1 + 4r^{}_{ji} + r^2_{ji})\right]\right\} \; , \label{eq:GammaMajpresG} \\
		\Gamma^{\rm M}_{\pm\mp,ij} &=& \frac{g^2_{ij}m^{}_i}{4\pi} \sqrt{1 - \beta^2_i} \left\{ \frac{1}{2} \left(1 - r^2_{ji}\right) (1 - r^{}_{ji})^2 + r^{}_{ji} (1 - r^2_{ji}) + 2r^2_{ji} \ln r^{}_{ji} \right. \nonumber \\
		&~& \left. + (\beta^{-2}_i - 1) \left[ (1+r^{}_{ji}+r^2_{ji}) r^{}_{ji} \ln r^{}_{ji} + \frac{1}{4} (1 - r^2_{ji}) (1 + 4r^{}_{ji} + r^2_{ji})\right] \right\} \; , \label{eq:GammaMajflipG}
	\end{eqnarray}
	for $\beta^{}_i > \beta^*_{ji} \equiv (1 - r^2_{ji})/(1 + r^2_{ji})$; and 
	\begin{eqnarray}
		\Gamma^{\rm M}_{\pm\pm,ij} &=& \frac{g^2_{ij}m^{}_i}{4\pi} \sqrt{1 - \beta^2_i} \left\{ \frac{1}{2} \left(1 - r^2_{ji}\right) (1 - r^{}_{ji})^2 - (1+r^{}_{ji}+r^2_{ji}) r^{}_{ji} \beta^{-1}_i \right. \nonumber \\
		&& \left. + \left[ r^{}_{ji} + \frac{1}{2} (\beta^{-2}_i - 1) (1+r^{}_{ji}+r^2_{ji})\right]  r^{}_{ji} \ln \left(\frac{1+\beta^{}_i}{1-\beta^{}_i}\right) \right\} \; , \label{eq:GammaMajpresL} \\
		\Gamma^{\rm M}_{\pm\mp,ij} &=& \frac{g^2_{ij}m^{}_i}{4\pi} \sqrt{1 - \beta^2_i} \left\{ \frac{1}{2} \left(1 - r^2_{ji}\right) (1 - r^{}_{ji})^2 + (1+r^{}_{ji}+r^2_{ji}) r^{}_{ji} \beta^{-1}_i \right. \nonumber \\
		&& \left. - \left[ r^{}_{ji} + \frac{1}{2} (\beta^{-2}_i - 1) (1+r^{}_{ji}+r^2_{ji})\right]  r^{}_{ji} \ln \left(\frac{1+\beta^{}_i}{1-\beta^{}_i}\right) \right\} \; , \label{eq:GammaMajflipL}
	\end{eqnarray}
	for $\beta^{}_i \leqslant \beta^*_{ji} \equiv (1 - r^2_{ji})/(1 + r^2_{ji})$. Some helpful comments on the decay rates in Eqs.~(\ref{eq:GammaMajpresG})-(\ref{eq:GammaMajflipL}) are in order.

	\begin{figure}[t]
		\centering
		\includegraphics[scale=1.0]{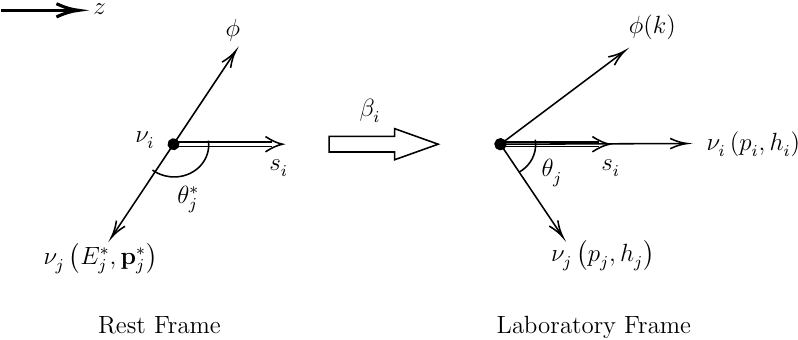}
		\caption{The sketch of the neutrino decay $\nu_i^{}(p_i^{},h_i^{}) \to \nu_j^{}(p_j^{},h_j^{}) + \phi(k)$ in the RF (left) and the LF (right), where the kinematic quantities are related by the Lorentz boost $\beta_i^{}$ on $\nu_i^{}$ along the positive direction of $z$-axis.}
		\label{fig:sketch}
	\end{figure}

	\begin{itemize}
		\item First of all, we should note that $\beta^*_{ji} \equiv (1 - r^2_{ji})/(1 + r^2_{ji})$ is just the velocity of the daughter neutrino $\nu^{}_j$ in the rest frame (RF) of the parent neutrino $\nu^{}_i$, where one can easily find the energy and momentum of $\nu^{}_j$ as $E^*_j = m_i^{}(1 + r_{ji}^2)/2$ and $|{\bf p}^*_j| = m_i^{}(1 - r_{ji}^2)/2$. The kinematic quantities with the superscript ``$*$" refer to those in the RF of the parent neutrino. In Fig.~\ref{fig:sketch}, we draw a sketch of the neutrino decay in the RF and the LF, on which the relevant kinematics quantities are marked. The decay in the LF can be treated as being obtained by performing a Lorentz boost on $\nu_i^{}$ along the positive direction of the $z$-axis, which is also the spin direction. By taking the limit of $\beta^{}_i \to 0$ in Eqs.~(\ref{eq:GammaMajpresL}) and (\ref{eq:GammaMajflipL}), we can derive the decay rates in the RF, i.e.,
		\begin{eqnarray}
			\Gamma^{\rm M *}_{\pm\pm,ij} = \Gamma^{\rm M *}_{\pm \mp,ij} = \frac{g^2_{ij} m^{}_i}{8\pi} \left(1 - r^2_{ji}\right) (1 - r^{}_{ji})^2 \; . \label{eq:GammaRF}
		\end{eqnarray} 
		
		\item It is worth mentioning that the total decay rates of $\nu^{}_i(p^{}_i, h^{}_i) \to \nu^{}_j(p^{}_j, h^{}_j) + \phi (k)$ summed over both helical states of $\nu^{}_j$ in the LF can be read off from Eqs.~(\ref{eq:GammaMajpresG})-(\ref{eq:GammaMajflipL}), namely,
		\begin{eqnarray} \label{eq:GammaLFtotal}
			\Gamma^{\rm M}_{\pm,ij} \equiv \Gamma^{\rm M}_{\pm\pm,ij} + \Gamma^{\rm M}_{\pm \mp,ij} = \frac{g^2_{ij} m^{}_i}{4\pi} \sqrt{1 - \beta^2_i} (1 - r^2_{ji}) (1 - r^{}_{ji})^2 \; ,
		\end{eqnarray}
		which are related to their counterparts $\Gamma^{{\rm M} *}_{\pm,ij} \equiv \Gamma^{{\rm M} *}_{\pm\pm,ij} + \Gamma^{{\rm M} *}_{\pm \mp,ij}$ in the RF expressed in Eq.~(\ref{eq:GammaRF}) by the Lorentz factor. For later convenience, the total decay rate of $\nu_i^{}(p_i^{},h_i^{})$ will be further defined as $\Gamma_{\pm,i}^{\rm M} \equiv \sum_j \Gamma^{\rm M}_{\pm,ij}$ by summing over all the possible decay channels with daughter neutrinos $\nu_j^{}(p_j^{},h_j^{})$. Except for the Lorentz factor, the total decay rates $\Gamma^{\rm M}_{\pm,ij}$ are independent of the velocity $\beta^{}_i$ of the parent neutrino $\nu^{}_i$. However, the rates of $\nu^{}_i(p^{}_i, h^{}_i)$ decays into different helical states $\nu^{}_j(p^{}_j, h^{}_j)$ indeed depend on the velocity $\beta^{}_i$ of the parent neutrino and on whether $\beta^{}_i$ is larger or smaller than the velocity $\beta^*_{ji}$.

		\item In the RF, the angular distribution of daughter neutrinos is isotropic, and their directions are described by the angle $\theta_j^*$ between the three-momentum ${\bf p}_j^*$ and the $z$-axis. In the LF, the longitudinal momentum of $\nu_j^{}$ parallel to $z$-axis, i.e., $p_{j,z}^{}$, is related to the energy and three-momentum in the RF by
		\begin{eqnarray}
			p_{j,z}^{} \equiv |{\bf p}_j^{}|\cos\theta_j^{} = \gamma_i^{} \left(\beta_i^{} E_{j}^* + |{\bf p}_j^{*}|\cos\theta_j^*\right) \;,
		\end{eqnarray}
		with the Lorentz factor $\gamma^{}_i \equiv 1/\sqrt{1 - \beta^2_i}$. There exists a critical value $\beta^{*}_{ji} \equiv (1 - r^2_{ji})/(1 + r^2_{ji})$ for $\beta_i^{}$, which is determined by whether the direction of the daughter neutrino momentum is perpendicular to the boost direction, i.e., $p_{j,z}^{}=0$ or equivalently $\theta_j^{} = \pi/2$. Therefore, the counterpart $\widehat{\theta}_j^*$ in the RF satisfies $\cos\widehat{\theta}_j^* = - \beta_i^{} / \beta_{ji}^*$. In the case of $\beta_i^{} < \beta_{ji}^*$, daughter neutrinos with $\theta_j^* < \widehat{\theta}_j^*$ in the RF are all boosted to the LF with $\theta_j^{} < \pi/2$, while those with $\theta_j^* > \widehat{\theta}_j^*$ to the LF with $\theta_j^{} > \pi/2$. On the other hand, for $\beta_i^{} > \beta_{ji}^*$, all the daughter neutrinos are boosted to $ \theta_j^{} < \pi/2 $, and there is no backward emission. The dependence on the velocity $\beta_i^{}$ reflects these two different configurations of angular distributions.
		
		\item With a small but nonzero velocity $\beta^{}_i \ll 1$, the total decay rates in Eqs.~(\ref{eq:GammaMajpresL}) and (\ref{eq:GammaMajflipL}) approximate to
		\begin{eqnarray}
			\Gamma^{\rm M}_{\pm\pm,ij} &\approx& \frac{g^2_{ij}m^{}_i}{4\pi} \left[ \frac{1}{2} (1-r^2_{ji}) (1 - r^{}_{ji})^2 - \frac{2}{3}  r^{}_{ji} \beta^{}_i \left(1-r^{}_{ji}\right)^2 \right] \; , \\
			\Gamma^{\rm M}_{\pm\mp,ij} &\approx& \frac{g^2_{ij}m^{}_i}{4\pi} \left[ \frac{1}{2} (1-r^2_{ji}) (1 - r^{}_{ji})^2 + \frac{2}{3}  r^{}_{ji} \beta^{}_i \left(1-r^{}_{ji}\right)^2 \right] \; ,
		\end{eqnarray}
		where only the first-order corrections of ${\cal O}(\beta^{}_i)$ are maintained. One can recognize that the rates of helicity-preserving and -changing decays are different for non-relativistic neutrinos, but the difference is proportional to the velocity $\beta^{}_i$. On the other hand, for ultra-relativistic neutrinos with $|{\bf p}^{}_i| \gg m^{}_i$, we have $\beta^{}_i \to 1$ and thus the formulae in Eqs.~(\ref{eq:GammaMajpresG}) and (\ref{eq:GammaMajflipG}) should be used. In this limit, the decay rates will be generally suppressed by the inverse of the Lorentz factor $\gamma^{-1}_i$, so the contributions proportional to $(\beta^{-2}_i - 1)$ can be safely neglected. In this case, we obtain
		\begin{eqnarray}
			\Gamma^{\rm M}_{\pm\pm,ij} &\approx& \frac{g^2_{ij} m^{}_i}{4\pi} \sqrt{1 - \beta^2_i} \left[ \frac{1}{2} (1 - r^2_{ji}) (1 - r^{}_{ji})^2  - r^{}_{ji} (1 - r^2_{ji}) - 2r^2_{ji} \ln r^{}_{ji} \right] \; , \\
			\Gamma^{\rm M}_{\pm\mp,ij} &\approx& \frac{g^2_{ij} m^{}_i}{4\pi} \sqrt{1 - \beta^2_i} \left[ \frac{1}{2} (1 - r^2_{ji}) (1 - r^{}_{ji})^2  + r^{}_{ji} (1 - r^2_{ji}) + 2r^2_{ji} \ln r^{}_{ji} \right] \; ,
		\end{eqnarray}
		which are exactly the same results previously derived, i.e., Eqs.~(10a) and (10b) in Ref.~\cite{Funcke:2019grs}. Except for the overall factor $\sqrt{1 - \beta^2_{i}}$, the difference between the helicity-preserving and -changing decay rates is only determined by the neutrino mass ratio $r^{}_{ji}$.
\end{itemize}

\subsubsection{Rate Asymmetries}
	
Thus far, we have not considered the neutrino mass spectrum, which is of crucial importance for the determination of the mass ratio $r^{}_{ji}$ or the velocity $\beta^*_{ji}$ in the RF. According to the latest global analysis of neutrino oscillation data~\cite{Gonzalez-Garcia:2021dve}, one gets $\Delta m^2_{21} \equiv m^2_2 - m^2_1 \approx 7.4\times 10^{-5}~{\rm eV}^2$ and $|\Delta m^2_{31}| \equiv |m^2_3 - m^2_1| \approx 2.5\times 10^{-3}~{\rm eV}^2$. For definiteness, we assume that the lightest neutrino is rather light, namely, $m^{}_1 \ll m^{}_2 < m^{}_3$ in the case of NO and $m^{}_3 \ll m^{}_1 < m^{}_2$ in the case of IO. For illustration, $m^{}_1 (m^{}_3) = 0.1~{\rm meV}$ will be taken in the NO (IO) case. Therefore, it is straightforward to obtain
\begin{eqnarray}
	m^{}_1 = 0.1~{\rm meV} \;, \quad m^{}_2 \approx 8.6~{\rm meV} \;, \quad m^{}_3 \approx 50~{\rm meV} \; ,
\end{eqnarray}
for the NO case, and
\begin{eqnarray}
	m^{}_3 = 0.1~{\rm meV} \;, \quad m^{}_1 \approx 50~{\rm meV} \;, \quad m^{}_2 \approx 50.7~{\rm meV} \; ,
\end{eqnarray}
for the IO case. In the NO case, neutrino decays proceed in three different channels $\nu^{}_3 \to \nu^{}_2 + \phi$, $\nu^{}_3 \to \nu^{}_1 + \phi$ and $\nu^{}_2 \to \nu^{}_1 + \phi$, so the relevant mass ratios are 
 \begin{eqnarray}
     r^{}_{23} \approx 0.17\;, \quad r^{}_{13} \approx 0.002 \;, \quad  r^{}_{12} \approx 0.01 \;.
 \end{eqnarray}
 In the IO case, we also have three decay channels $\nu^{}_2 \to \nu^{}_1 + \phi$, $\nu^{}_2 \to \nu^{}_3 + \phi$ and $\nu^{}_1 \to \nu^{}_3 + \phi$ with the mass ratios 
 \begin{eqnarray}
     r^{}_{12} \approx 0.99\;, \quad r^{}_{32} \approx r^{}_{31} \approx 0.002\;.
 \end{eqnarray}
 Given these values of $r^{}_{ji}$, the difference between the helicity-preserving decay rates $\Gamma^{\rm M}_{\pm\pm,ij}$ and the helicity-changing ones $\Gamma^{\rm M}_{\pm\mp,ij}$ is actually sensitive to the velocity $\beta^{}_i$ of the parent neutrino $\nu^{}_i$ in the LF, as we have explained. 

 \begin{figure}[t]
		\centering
		\includegraphics[scale=0.54]{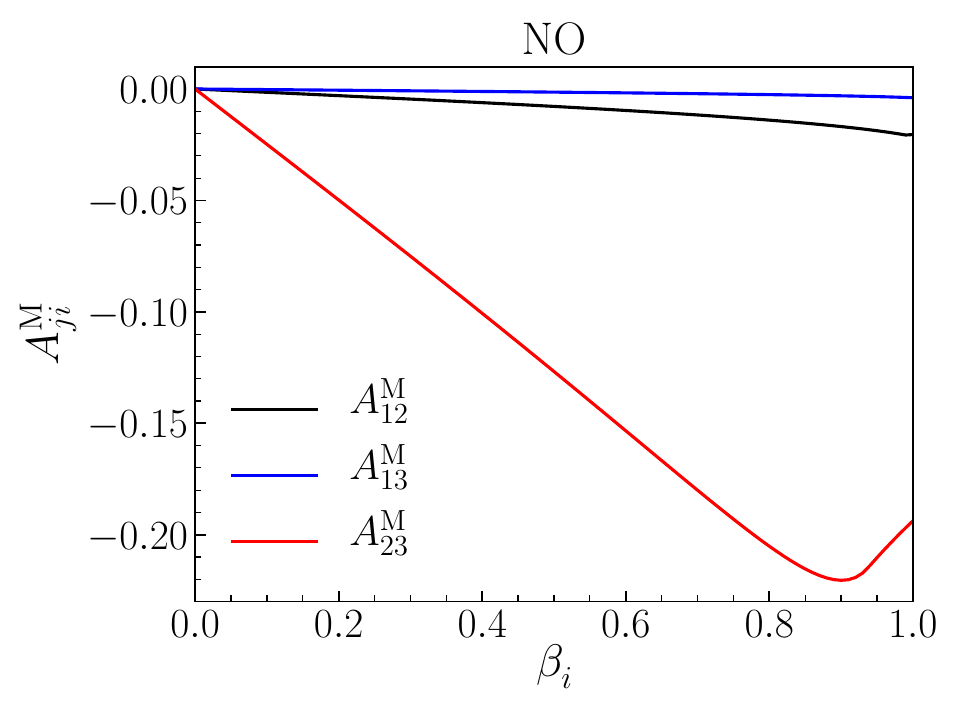} \hspace{-0.5cm}
		\includegraphics[scale=0.54]{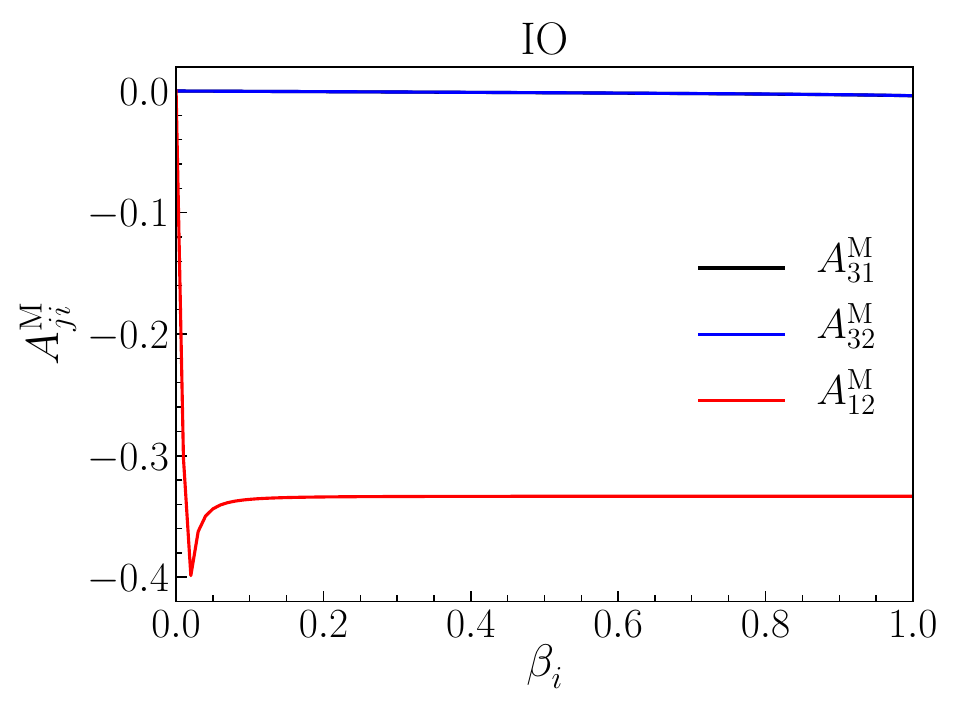}
		\vspace{-0.5cm}
		\caption{The asymmetry between the helicity-preserving and -changing decay rates $A^{\rm M}_{ji} $ for the channel $\nu^{}_i \to \nu^{}_j + \phi $ versus the velocity of the parent neutrino $\beta^{}_i$. In the NO case (left panel), the decay-rate asymmetry is displayed as black, blue and red curves with $ji = 12,13,23$, respectively. In the IO case (right panel), the asymmetry is accordingly shown as black, blue and red curves with $ji = 31,32,12$.}
		\label{fig:A}
	\end{figure}
	
To have a more direct feeling about the difference between two decay channels, we introduce the asymmetry between the helicity-preserving and -changing decay rates as
\begin{eqnarray}
	A^{\rm M}_{ji} &\equiv& \frac{\Gamma^{\rm M}_{\pm\pm,ij} - \Gamma^{\rm M}_{\pm\mp,ij}}{\Gamma^{\rm M}_{\pm\pm,ij} + \Gamma^{\rm M}_{\pm\mp,ij}} \; .
	\label{eq:AMji}
\end{eqnarray}
With the help of Eqs.~(\ref{eq:GammaMajpresG})-(\ref{eq:GammaMajflipL}), we can find 
\begin{eqnarray}\displaystyle
		A^{\rm M}_{ji} &=& \frac{r^{}_{ji} \left(1 + r^{}_{ji} + r^2_{ji}\right)}{\beta_i^2 (1-r_{ji}^{})^2 (1-r_{ji}^{2})}\left[\ln \left(\displaystyle \frac{1+\beta_i^{}}{1-\beta_i^{} }\right) - 2 \beta_i^{}\right] - \frac{r^{}_{ji} \left(1 - r^{}_{ji} + r^2_{ji}\right)}{ (1-r_{ji}^{})^2 (1-r_{ji}^{2})} \ln \left(\displaystyle \frac{1+\beta_i^{}}{1-\beta_i^{} }\right)    \; , 
		\label{eq:AMjiL}
\end{eqnarray}
for $\beta_i^{} < \beta_{ji}^*$, which gives rise to $A^{\rm M}_{ji} \to 0 $ with $\ln\left[(1+\beta^{}_i)/(1 - \beta^{}_i)\right] = 2\beta^{}_i + 2\beta^3_i/3 + {\cal O}(\beta^5_i)$ in the limit of $\beta^{}_i \to 0$; and 
\begin{eqnarray}
	A^{\rm M}_{ji} &=& -\frac{2 r^{}_{ji} \left[(1-\beta^2_i)(1+r^2_{ij}) + (1+\beta^2_i)r^{}_{ji}\right] \ln r^{}_{ji}}{\beta_i^2 (1-r_{ji}^{})^2 (1 - r_{ji}^{2})}  - \frac{\left(1- \beta_i^2\right) \left(1+r_{ji}^2\right) + 4 r_{ji}^{}}{2\beta^2_i (1-r^{}_{ji})^2}   \; ,
	\label{eq:AMjiG}
\end{eqnarray}
for $\beta_i^{} > \beta_{ji}^*$. In the limit of $\beta^{}_i \to 1$, we obtain 
\begin{eqnarray} \label{eq:AMjilimit}
	A^{\rm M}_{ji} \to - \frac{2 r_{ji}^{} \left(1 + 2 r_{ji}^{} \ln r_{ji}^{} - r_{ji}^2\right)}{(1-r_{ji}^{})^2 (1 - r_{ji}^{2})} \; ,
\end{eqnarray}
from Eq.~(\ref{eq:AMjiG}). The asymmetry $A^{\rm M}_{ji}$ has been plotted as the function of $\beta^{}_i$ in Fig.~\ref{fig:A} in the NO case (left panel) and the IO case (right panel). Some comments on the rate asymmetry can be found below.
\begin{itemize}
	\item In the NO case, as $r^{}_{13} \approx 0.002$ and $r^{}_{12} \approx 0.01$ are extremely small, one can get $\beta^*_{13} \approx \beta^*_{12} \approx 1$. Therefore, for $\nu^{}_2 \to \nu^{}_1 + \phi$ and $\nu^{}_3 \to \nu^{}_1 + \phi$ decays, the final-state $\nu^{}_1$ is ultra-relativistic in the RF and the formulae in Eqs.~(\ref{eq:GammaMajpresL}) and (\ref{eq:GammaMajflipL}) are valid for most values of $\beta^{}_i$ except for $\beta^{}_i \to 1$. As the asymmetries $A^{\rm M}_{13}$ (blue curve) and $A^{\rm M}_{12}$ (black curve) are suppressed by the small values of $r^{}_{13}$ and $r^{}_{12}$, respectively, there is no significant difference between the rates of helicity-preserving and -changing decays. On the other hand, for the decay channel $\nu^{}_3 \to \nu^{}_2 + \phi$, we have $r^{}_{23} \approx 0.17$ and $\beta^*_{23} \approx 0.944$. As the value of $r^{}_{23}$ is not that small, we should expect a more significant asymmetry $A^{\rm M}_{23}$ (red curve) between two decay rates. For ultra-relativistic $\nu_3^{}$ with $\beta_3^{} \to 1$, one can estimate the asymmetry $A_{23}^{\rm M} \approx -0.18$ from Eq.~(\ref{eq:AMjilimit}), as also indicated in the left panel of Fig.~\ref{fig:A}. Especially, for the critical velocity $\beta^{}_3 = \beta^*_{23}$, we have $A^{\rm M}_{23} \approx -0.21 $, which represents an asymmetry of around $20\%$ between two decay rates. This difference indicates that for the decay $\nu^{}_3 \to \nu^{}_2 +\phi$, there are a significant number of daughter neutrinos with opposite helicities to those of high-velocity parent neutrinos. 
		
	\item In the IO case, as $r^{}_{31} \approx r^{}_{32} \approx 0.002$, we have $\beta^{*}_{31} \approx \beta^{*}_{32} \approx 0.99$. Therefore, for $\nu^{}_1 \to \nu^{}_3 +\phi $ and $\nu^{}_2 \to \nu^{}_3 + \phi $, the final-state $\nu^{}_3$ is ultra-relativistic in the RF and the formulae in Eqs.~(\ref{eq:GammaMajpresL}) and (\ref{eq:GammaMajflipL}) could perfectly describe the decay rates except for $\beta^{}_i \to 1$. Furthermore, with small mass ratios, the asymmetries $A_{31}^{\rm M}$ (black curve) and $A_{32}^{\rm M}$ (blue curve) are also highly suppressed. On the other hand, for the decay channel $\nu^{}_2 \to \nu^{}_1 + \phi$ with $\beta^{*}_{12} \approx 0.02 $, the decay rate is mostly described by expressions in Eqs.~(\ref{eq:GammaMajpresG}) and (\ref{eq:GammaMajflipG}). The mass ratio $r^{}_{12} \approx 0.99$ suggests $A_{12}^{\rm M} \approx -0.33$ when $\beta_2^{} \to 1$ (red curve), and implies a large asymmetry of at most 40\%, which further demonstrates the necessity of considering the helicity-changing decays. 
\end{itemize}
	
Those rate asymmetries will strongly influence the helicity distribution of daughter particles, thereby changing the capture rates of C$\nu$B. However, before the discussion of background neutrino decays, we pay particular attention to the observational constraints on neutrino lifetimes.
	
\subsection{Observational Constraints}
	
The visible and invisible decays of massive neutrinos have been studied extensively in the literature. For visible decays, in which a heavy neutrino decays into a lighter one and a photon $\nu_i^{} \to \nu_j^{} + \gamma$, the decay rate can be written as~\cite{Xing:2011zza}
\begin{eqnarray}
	\Gamma(\nu_i^{} \to \nu_j^{} + \gamma) = 5.3~{\rm s}^{-1} \left(1-r_{ji}^2\right)^3 \left(\frac{m_i^{}}{1~{\rm eV}}\right)^3 \left(\frac{\mu_{\rm eff}^{}}{\mu_{\rm B}^{}}\right)^2 \;,
\end{eqnarray}
where $\mu_{\rm eff}^{}$ is the effective magnetic dipole moment. In the minimal extension of the SM with a finite neutrino mass term, the magnetic dipole moment is as small as $\mu_{\rm eff}
^{} \approx 10^{-23}\mu_{\rm B}^{}$ with the Bohr magneton $\mu_{\rm B}^{} = e/(2m_e^{}) \approx 0.3~{\rm MeV}^{-1}$ in natural units~\cite{ParticleDataGroup:2022pth}. Therefore, the lifetime of unstable neutrinos can be as long as $\tau_\nu^{} \approx 10^{49}~{\rm s}$, much longer than the age of our Universe $t_0^{} \approx 4 \times 10^{17}~{\rm s}$. Such a long lifetime satisfies all observational constraints from the precise measurement on distortions of the cosmic microwave background (CMB) spectrum~\cite{Mirizzi:2007jd,Aalberts:2018obr}, the redshifted $21~{\rm cm}$ line of neutral hydrogen atoms~\cite{Chianese:2018luo}, and the solar $X$- and gamma-rays~\cite{Raffelt:1985rj}. 

In this work, we focus on the invisible decays of massive neutrinos, which are also subject to various observational constraints~\cite{Kim:1990km, Funcke:2019grs, Palomares-Ruiz:2005zbh, Hannestad:2005ex, Pasquini:2015fjv, BahaBalantekin:2018ppj, Balantekin:2018ukw, deGouvea:2019qre, Abdullahi:2020rge, Cheng:2020rla, deGouvea:2021ual, deGouvea:2022cmo, Huang:2022udc, Batell:2024hzo}. To simplify the discussions, we further set all the couplings to be equal, i.e., $g_{ij}^{} \equiv g$. Given the neutrino mass spectrum, the coupling constant $g$ becomes the only free parameter in the scenarios under discussions. In general, the existence of a massless pseudo-scalar $\phi$ with a sizable coupling to neutrinos could significantly change the thermal history of the Universe. For large enough couplings, $\phi$ will be abundantly produced in the early Universe and reach thermal equilibrium through both decays $\nu^{}_i \to \nu^{}_j + \phi$ and two-body scatterings $\nu^{}_i + \nu^{}_j \to \phi + \phi$. Therefore, the massless $\phi$ serves as an extra radiation that greatly affects the Big Bang Nucleosynthesis (BBN) and the CMB power spectrum. It is obvious that the coupling will receive a stringent constraint from observations. Among all the observational constraints, the cosmological one is the most restrictive. For an ultra-relativistic parent neutrino of mass $m_\nu^{}$ decaying into a massless neutrino and an invisible scalar, the lower bound on the neutrino lifetime in the RF is~\cite{Barenboim:2020vrr}
\begin{eqnarray}
	\tau_0^{} \gtrsim 4 \times 10^{(5\cdots 6)}~{\rm s}\left(\frac{m_\nu^{}}{50~{\rm meV}}\right)^5 \;.
\end{eqnarray}
When converting this bound to that on the coupling constant, we have 
\begin{eqnarray}
	g \lesssim (1.6\cdots 5.0) \times 10^{-10}~({\rm for\ \nu_3^{}})\;, \quad g \lesssim (0.4 \cdots 1.3) \times 10^{-7}~({\rm for\ \nu_2^{}})\; ,
\end{eqnarray}
in the NO case. In the IO case, we get $g \lesssim (2 \cdots 6) \times 10^{-10}$ for both $\nu_1^{}$ and $\nu_2^{}$, since they are almost degenerate in mass. If the masses of daughter neutrinos are taken into account, a phase-space factor comes into play and the constraint on the lifetime will be weakened~\cite{Chen:2022idm}. The revised bound on the coupling constant reads
\begin{eqnarray}
	g \lesssim 3.2 \times 10^{-10}~({\rm for\ \nu_3^{}})\;, \quad g \lesssim 5.0 \times 10^{-8}~({\rm for\ \nu_2^{}})\;
\end{eqnarray} 
in the NO case, while $g \lesssim 4 \times 10^{-10}$ in the IO case.
	
Other constraints on the invisible neutrino decays come from the terrestrial experiments and astrophysical observations. The interaction described in Eq.~(\ref{eq:LagM}) gives rise to the neutrinoless double-beta decays with an extra scalar $\phi$. The non-observation of such a signal in the EXO-200 experiment provides a constraint on the coupling constant $g \lesssim (0.4 \cdots 0.9) \times 10^{-5}$~\cite{Kharusi:2021jez}. On the other hand, astrophysical constraints on the coupling can also be derived from BBN~\cite{Ahlgren:2013wba,Huang:2017egl,Escudero:2019gvw,Venzor:2020ova}, Supernova 1987A~\cite{Kamiokande-II:1987idp,Bionta:1987qt,Kolb:1987qy,Alekseev:1988gp,Farzan:2002wx,Zhou:2011rc,Shalgar:2019rqe,Fiorillo:2022cdq,Fiorillo:2023ytr,Akita:2023iwq,Martinez-Mirave:2024hfd}, solar neutrinos~\cite{Bahcall:1972my,Berezhiani:1991vk,Cleveland:1998nv,Beacom:2002cb,SAGE:2009eeu,Bellini:2011rx,KamLAND:2011fld,KamLAND:2014gul,Super-Kamiokande:2016yck,Borexino:2017uhp,SNO:2018pvg,Huang:2018nxj}, atmospheric and long-baseline accelerator neutrinos~\cite{Lipari:1999vh,Fogli:1999qt,Gonzalez-Garcia:2008mgl,Gomes:2014yua,Choubey:2018cfz}, high-energy astrophysical neutrinos~\cite{Ng:2014pca,Shoemaker:2015qul,Denton:2018aml,deSalas:2018kri,Song:2020nfh,Valera:2024buc}.
	
To avoid the existing constraint from cosmological observations, we take the coupling constant in the range $10^{-16} \lesssim g \lesssim 10^{-10}$ in this work, which is also allowed by other experimental tests. In fact, for $g \lesssim 10^{-7}$, the massless pesudo-scalar $\phi$ will never be in thermal equilibrium via the two-body scattering before the CMB formation, as demonstrated in Refs.~\cite{Hannestad:2005ex,Barenboim:2020vrr}. A benchmark value of $g = 10^{-12}$ will be chosen for illustration. Smaller couplings will make the effect of neutrino decays more insignificant, whereas much larger ones have been excluded. More discussions about the restrictions on neutrino lifetimes and the coupling constant can be found in Ref.~\cite{Sandner:2023ptm} and the references therein.
		
\section{Cosmic Neutrino Background}
\label{sec:cnb}

\subsection{Invisible Decays of C$\nu$B}

In the standard cosmology, the decoupling temperature of ordinary neutrinos due to the inefficiency of their weak interactions is around $T \approx 1~{\rm MeV}$, corresponding to the redshift $z_{\rm d}^{} \approx 10^{10}$. After the decoupling, the three-momentum ${\bf p}(z)$ and the temperature $T_\nu^{}(z)$ of neutrinos evolve with the redshift according to~\cite{Langacker:1982ih,Weinberg:2008zzc,Zhang:2008if}
\begin{eqnarray}
	\label{eq:p_T_redshift}
	|{\bf p}(z)| = \frac{1+z}{1+z_{\rm d}^{}} |{\bf p}(z_{\rm d}^{})|\;,\quad T_\nu^{}(z) = \frac{1+z}{1+z_{\rm d}^{}} T_\nu^{}(z_{\rm d}^{})\;.
\end{eqnarray}
Meanwhile, the neutrino spectrum obeys the modified Fermi-Dirac (FD) distribution
\begin{eqnarray}\label{eq:FD}
	f_{\rm FD}^{} \left[|{\bf p}(z)|,T_\nu^{}(z)\right] = \frac{1}{\exp\left[|{\bf p}(z)|/T_\nu^{}(z)\right] + 1} \;, 
\end{eqnarray} 
where the neutrino mass can be neglected due to the suppression by a factor of $(1+z)/(1+z^{}_{\rm d})$. The corresponding neutrino number density is given by $\overline{n} (T^{}_\nu) = 3\zeta(3)T^3_\nu/(4\pi^2)$. At the present, the neutrino temperature is $T_\nu^{0} \equiv T^{}_\nu(z = 0) \approx 1.95~{\rm K} \approx 0.168~{\rm meV}$, and thus $\overline{n} (T_\nu^0) \approx 56~{\rm cm}^{-3}$ for each flavor. For three flavors of neutrinos and antineutrinos, the total number density is about $336~{\rm cm}^{-3}$. Since one can always translate the dependence on the temperature into that on the redshift with the help of Eq.~(\ref{eq:p_T_redshift}), we express the number density of $\nu_i^{}$ at the redshift $z$ as $\overline{n}_i^{}(z)$.

For later convenience, we define the modified number density of $\nu_i^{}$ as $n_i^{}(z) \equiv \overline{n}_i^{}(z) a(z)^3 $ with the scale factor $a(z) \equiv 1/(1+z)$. For absolutely stable neutrinos, the modified number density of $\nu_i^{}$ remains a constant, i.e., $n_0^{} \equiv n_i^{}(0) = n_i^{}(z^{}_{\rm d}) \approx 56~{\rm cm}^{-3}$, as the effect due to the expansion of the Universe on the number density has been absorbed into the scale factor. In consideration of the decays $\nu^{}_i \to \nu^{}_j +\phi$, the number densities of parent and daughter neutrinos with helicities $h =\pm 1$ will be redshift-dependent and denoted as $n_i^{\pm}(z)$ and $n_j^{\pm}(z)$, respectively. 

Strictly speaking, one must numerically solve the complete set of Boltzmann equations for the number densities of $\nu_i^{}$, $\nu_j^{}$ and $\phi$ in order to figure out their evolution behaviors. Here we adopt an intuitive and more efficient strategy, in which the number densities of parent and daughter neutrinos with the same helicity at the redshift $z$ read
\begin{eqnarray}\label{eq:daughter_number}
	n_i^{\pm} (z) = n_0^{} {\rm e}^{-\lambda_i^{} (z)}_{} \;, \quad n_j^{\pm} (z) = \left[n_0^{} - n_i^{\pm} (z) \right] {\cal B}^{\rm M+}_{ij} \;,
\end{eqnarray}
where ${\cal B}^{\rm M+}_{ij}\equiv \Gamma^{\rm M}_{\pm\pm,ij}/\Gamma^{\rm M}_{\pm,i}$ is the branching ratio of such a helicity-preserving decay channel. For the helicity-changing case, the branching ratio is ${\cal B}^{\rm M-}_{ij}\equiv \Gamma^{\rm M}_{\pm\mp,ij}/\Gamma^{\rm M}_{\pm,i}$. The explicit expressions of those decay rates are given in Eqs.~(\ref{eq:GammaMajpresG})-(\ref{eq:GammaMajflipL}), and the total decay rate $\Gamma^{\rm M}_{\pm,i} $ is defined below Eq.~(\ref{eq:GammaLFtotal}). Before explaining the suppression factor $\lambda_i^{} (z) $ in Eq.~(\ref{eq:daughter_number}), we need to first fix the characteristic redshift $z_i^{\rm decay}$ for $\nu_i^{}$ decaying at the rate of $\Gamma_{\pm,i}^{{\rm M}*}$ in the RF. To this end, we consider
\begin{eqnarray} 
	\label{eq:lambda_tilde}
	\widetilde{\lambda}_i^{} (z_i^{\rm decay}) \equiv \int^\infty_{z_i^{\rm decay}} \frac{{\rm d} z \ \Gamma_{\pm,i}^{{\rm M} *}}{(1+z) H(z) \gamma_i^{}(z)} \;,
\end{eqnarray}
with the Lorentz factor defined as
\begin{eqnarray} \label{eq:Lorentz_factor}
	\gamma^{}_i(z) \equiv E_i^{} (z)/m_i^{} = 1/\sqrt{1-\beta_i^{2}(z)} \;.
\end{eqnarray} 
Some explanations for the parameters $z^{\rm decay}_i$ and $\widetilde{\lambda}_i^{} (z_i^{\rm decay})$ are necessary. First, the velocity $\beta_i^{}(z)$ is related to the corresponding momentum ${\bf p}_i^{}(z)$ by $|{\bf p}_i^{}(z)| = m_i^{} \beta_i^{}(z)/\sqrt{1-\beta_i^{2}(z)}$. One should notice the redshift-dependence of the velocity and momentum arises from the FD distribution at different temperatures. The Hubble parameter $H(z)$ is
\begin{eqnarray} \label{eq:hubble}
	H(z) \equiv H_0^{} \sqrt{\Omega_{\rm m}^{} (1+z)^3 + \Omega_{\rm r}^{} (1+z)^4 + \Omega_\Lambda^{}}\;,
\end{eqnarray}
with the present-day value $H_0^{} \approx 67.4~{\rm km}~{\rm s}^{-1}~{\rm Mpc}^{-1}$~\cite{ParticleDataGroup:2022pth}. The energy-density fractions of matter, radiation and dark energy are given by $\Omega_{\rm m}^{} \approx 0.315$, $\Omega_{\rm r}^{} \approx 5 \times 10^{-5}$ and $\Omega_\Lambda^{} \approx 0.685$, respectively~\cite{ParticleDataGroup:2022pth}. Then, one can estimate $z_i^{\rm decay}$ from the following equation
\begin{eqnarray}\label{eq:solve_zidecay}
	\frac{\displaystyle \int_{0}^{\infty} {\rm e}^{-\widetilde{\lambda}_i^{} (z_i^{\rm decay})} \; f_{\rm FD}^{} \left[|{\bf p}|,T_\nu^{}(z_i^{\rm decay})\right] |{\bf p}|^2\ {\rm d}|{\bf p}|}{\displaystyle \int_{0}^{\infty} f_{\rm FD}^{} \left[|{\bf p}|,T_\nu^{}(z_i^{\rm decay})\right] |{\bf p}|^2\ {\rm d}|{\bf p}|} = {\rm e}^{-1}\;.
\end{eqnarray}

\begin{figure}[t]
	\centering
	\includegraphics[scale=0.54]{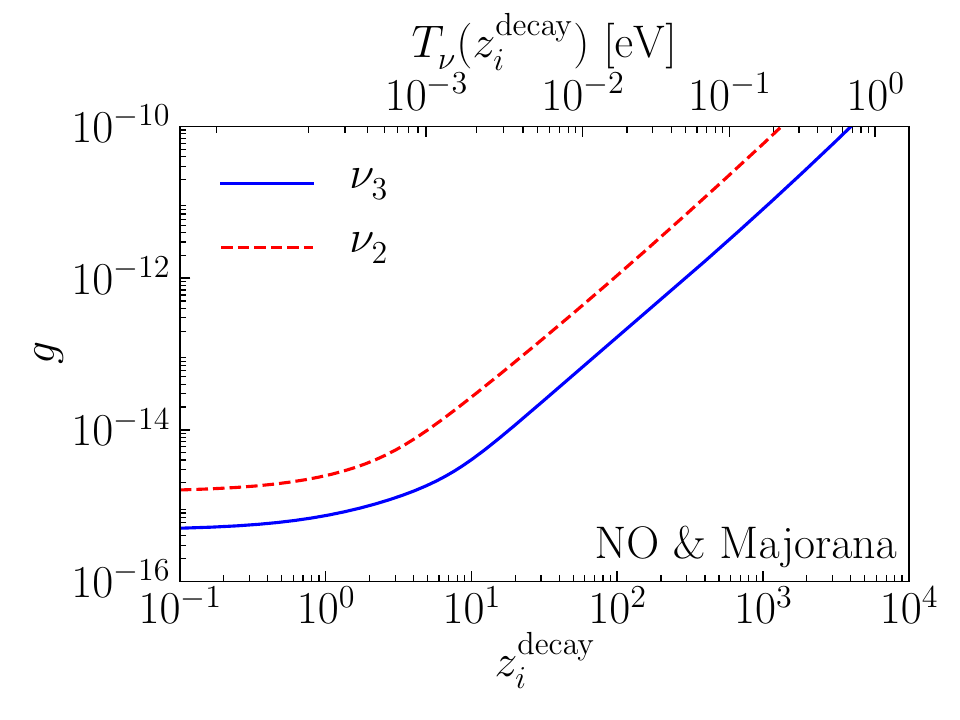} \hspace{-0.5cm}
	\includegraphics[scale=0.54]{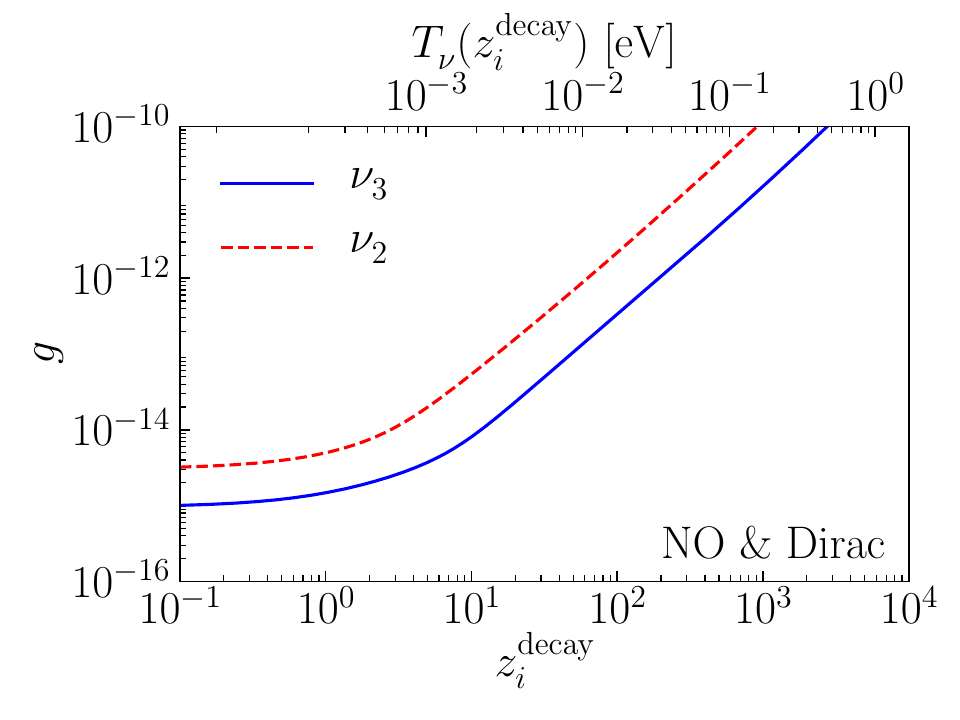}
	\includegraphics[scale=0.54]{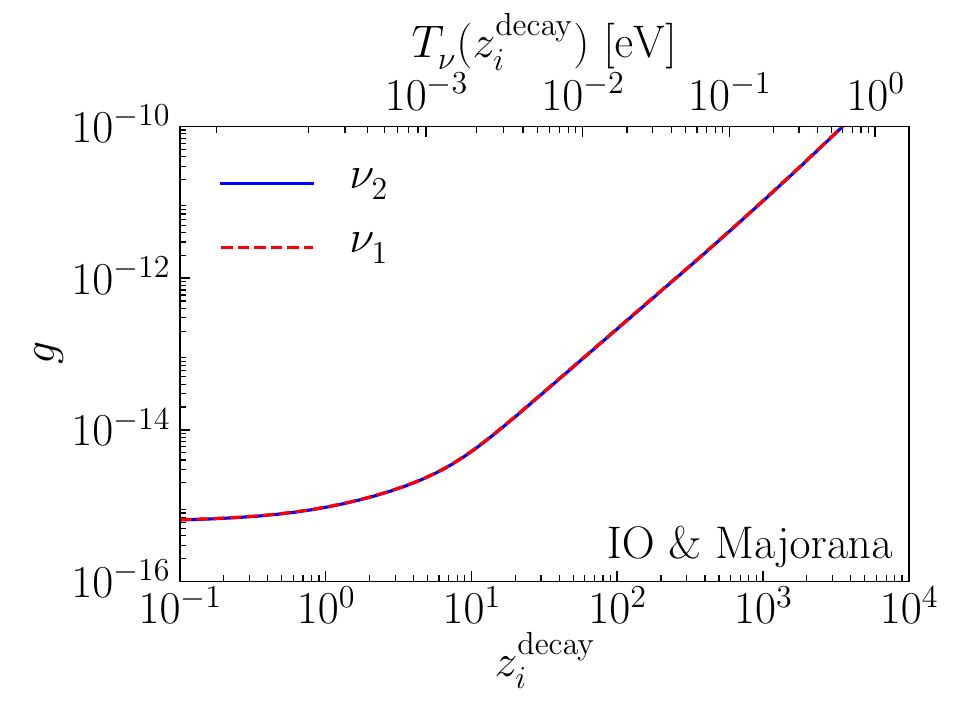}
	\hspace{-0.5cm}
	\includegraphics[scale=0.54]{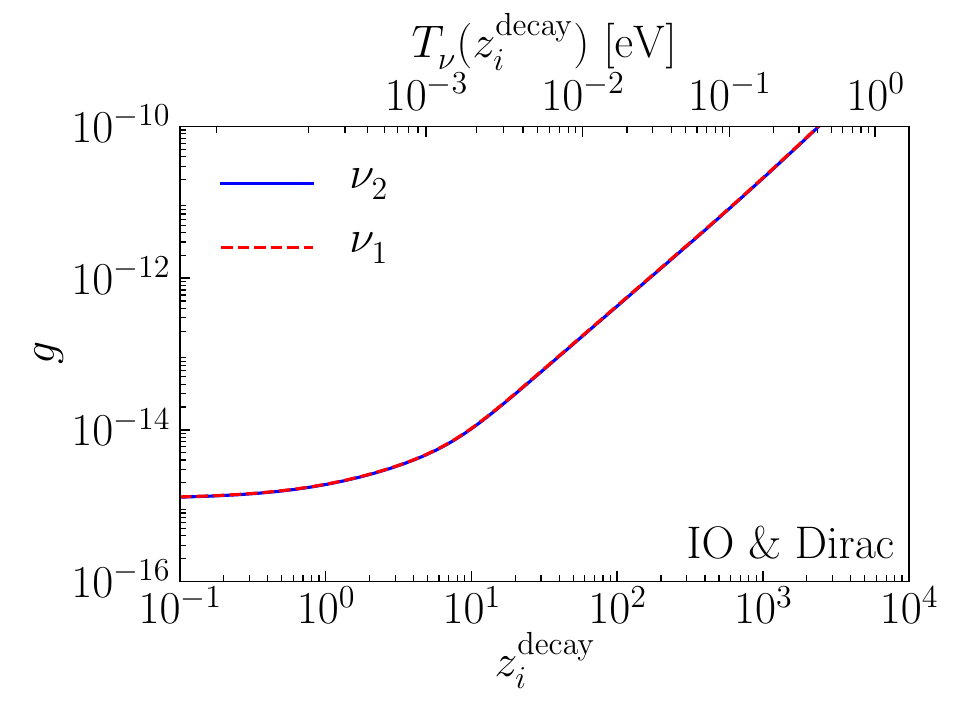} 
	\vspace{-0.5 cm}
	\caption{The relationship between the redshift of neutrino decays $z_i^{\rm decay} $ and the coupling constant $g$ in the case of NO (upper row) and IO (lower row), where neutrinos are assumed to be Majorana (left column) or Dirac particles (right column). The relation for the heaviest neutrino ($\nu_3^{}$ for NO or $\nu_2^{}$ for IO) is plotted in blue solid curves, while for the second-heaviest ($\nu_2^{}$ for NO or $\nu_1^{}$ for IO) it is shown in red dashed curves. The corresponding temperature $T^{}_\nu(z_i^{\rm decay})$ at the redshift $z_i^{\rm decay}$ is shown along the upper axes.}
	\label{fig:g_zdecay}
\end{figure}

From the definition of $z^{\rm decay}_i$ in Eq.~(\ref{eq:solve_zidecay}) and that of $\widetilde{\lambda}(z^{\rm decay}_i)$ in Eq.~(\ref{eq:lambda_tilde}), it is clear that $z^{\rm decay}_i$ characterizes the redshift when a substantial fraction of $\nu^{}_i$ starts to decay for a given coupling constant $g$ after taking account of its momentum distribution, whereas ${\rm e}^{-\widetilde{\lambda}(z^{\rm decay}_i)}$ stands for the suppression factor for the neutrino number density due to decays. The relationship between $z_i^{\rm decay}$ and the coupling constant $g$ is depicted in Fig.~\ref{fig:g_zdecay}. Generally speaking, we have $z^{\rm decay}_i$ growing with $g$. It is reasonable that a larger coupling constant leads to a shorter neutrino lifetime, implying that the decay happens at an earlier time with a larger redshift. Since the decay rate is proportional to the parent neutrino mass, as in Eqs.~(\ref{eq:GammaMajpresG})-(\ref{eq:GammaMajflipL}), we have $z_3^{\rm decay} > z_2^{\rm decay}$ in the case of NO (upper row) for the same coupling $g$, no matter whether neutrinos are Majorana (left panel) or Dirac (right panel) particles. However, as two unstable neutrinos in the IO case (lower row) are nearly degenerate in mass, the two curves for $\nu_2^{}$ and $\nu_1^{}$ almost overlap with each other. On the other hand, as the decay rate for Majorana neutrinos is four times larger than that in the Dirac case, the corresponding redshift is also relatively larger. Hence the physical meaning of $z^{\rm decay}_i$ is just a characteristic redshift when neutrinos decay.

Given the coupling constant in the range $10^{-14} \lesssim g \lesssim 10^{-10}$, one can notice from Fig.~\ref{fig:g_zdecay} that neutrinos decay around $T^{}_{\nu} \in [10^{-3}, 10^{-1}]~{\rm eV}$, which is far lower than the neutrino decoupling temperature. Therefore, it is reasonable to calculate the suppression factor for the neutrino number density by using~\cite{Baerwald:2012kc}
\begin{eqnarray} \label{eq:lambda}
	\lambda_i^{} (z) = \int_z^{z_i^{\rm decay}} \frac{{\rm d} z^\prime \ \Gamma_{\pm,i}^{{\rm M} *}}{(1+z^\prime) H(z^\prime) \gamma(z^\prime)} \;,
\end{eqnarray}
in terms of the redshift $z_i^{\rm decay}$. We emphasize the redshift-dependence in Eqs.~(\ref{eq:daughter_number}) and (\ref{eq:lambda}) is not the same as that in $\overline{n}_i^{}(z)$. The latter one comes just from the expansion of the Universe, while the former takes account of neutrino decays as well. 

The secondary decay from the second-heaviest neutrinos should also be taken into account, and the strategy is basically the same as before. We take the NO case as an example to explain the treatment of secondary decays of $\nu_2^{}$, which is the second-heaviest neutrino. The characteristic redshift $z_2^{\prime \rm decay}$ at which the secondary decay of $\nu_2^{}$ occurs can be similarly defined with the help of Eq.~(\ref{eq:solve_zidecay}). However, there are two main differences. First, the upper limit of the integral in Eq.~(\ref{eq:lambda_tilde}) to define $\widetilde{\lambda}_2^{}$ should be changed from infinity into $z_3^{\rm decay}$, below which the secondary $\nu^{}_2$ is produced from $\nu^{}_3$ decays. Second, the FD distribution $f_{\rm FD}^{}$ should not be adopted any more, but needs to be replaced by the distribution function of the produced $\nu_2^{}$ that will be discussed below. The suppression factor $\lambda_2^\prime (z)$, describing the evolution of the number density for $\nu^{}_2$, is similar to that in Eq.~(\ref{eq:lambda}) but now in terms of $z_2^{\prime \rm decay}$.

\begin{figure}[t!]
	\centering
	\includegraphics[scale=0.5]{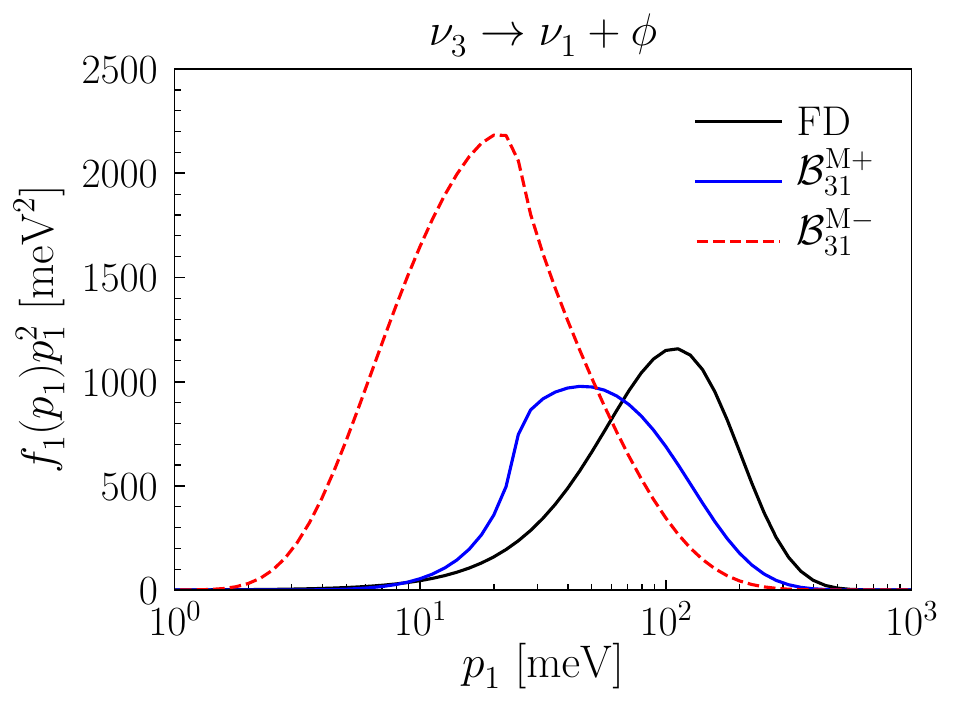}
	\includegraphics[scale=0.5]{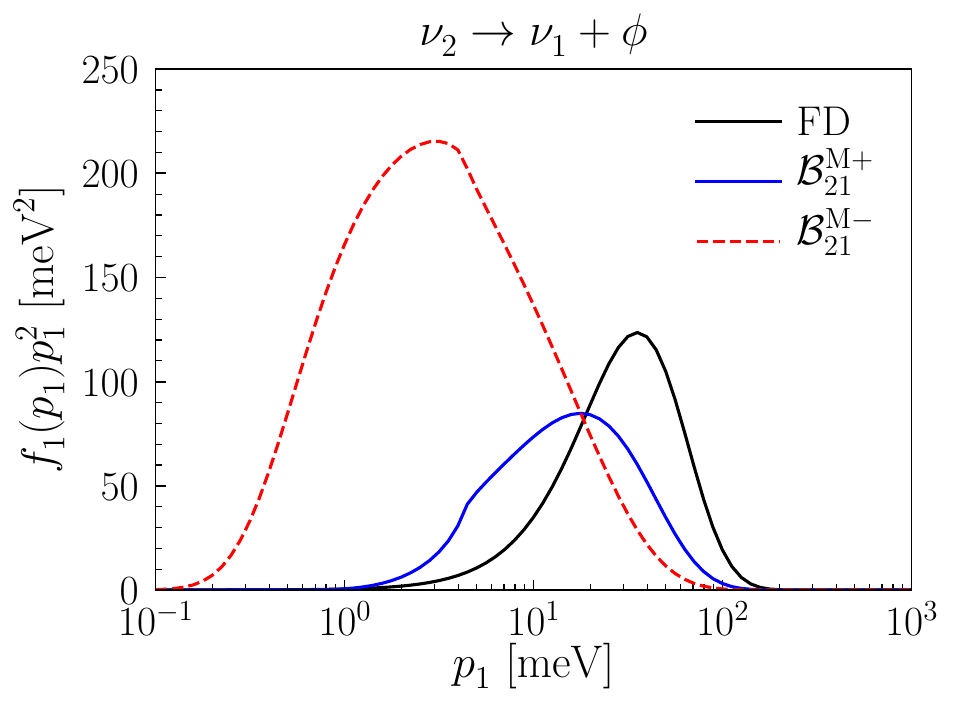}\\ \vspace{0.3cm}
	\includegraphics[scale=0.5]{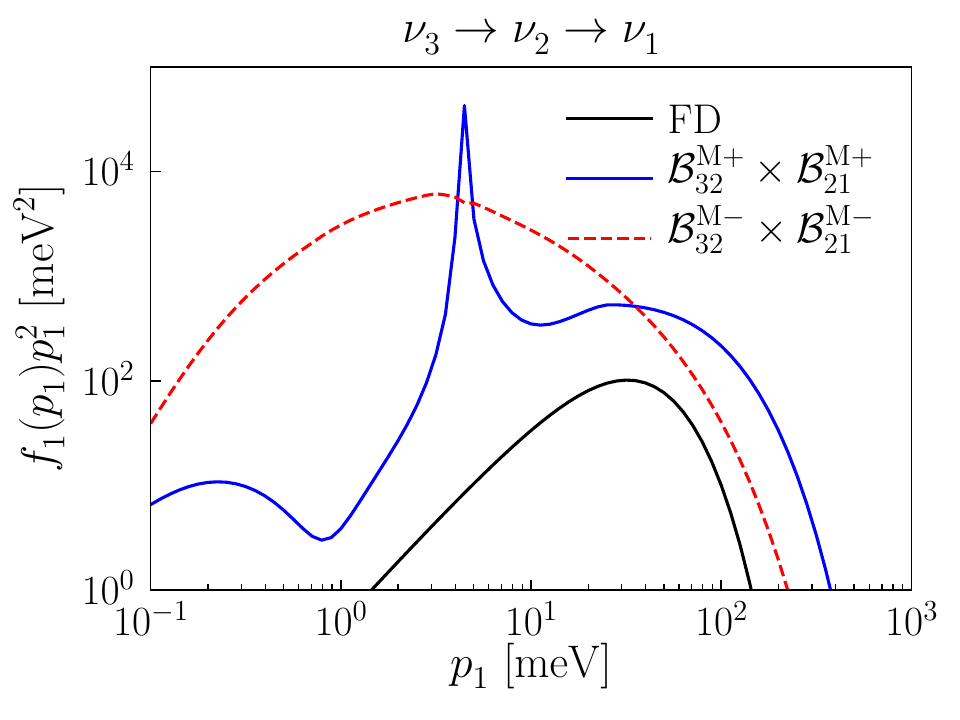}
	\includegraphics[scale=0.5]{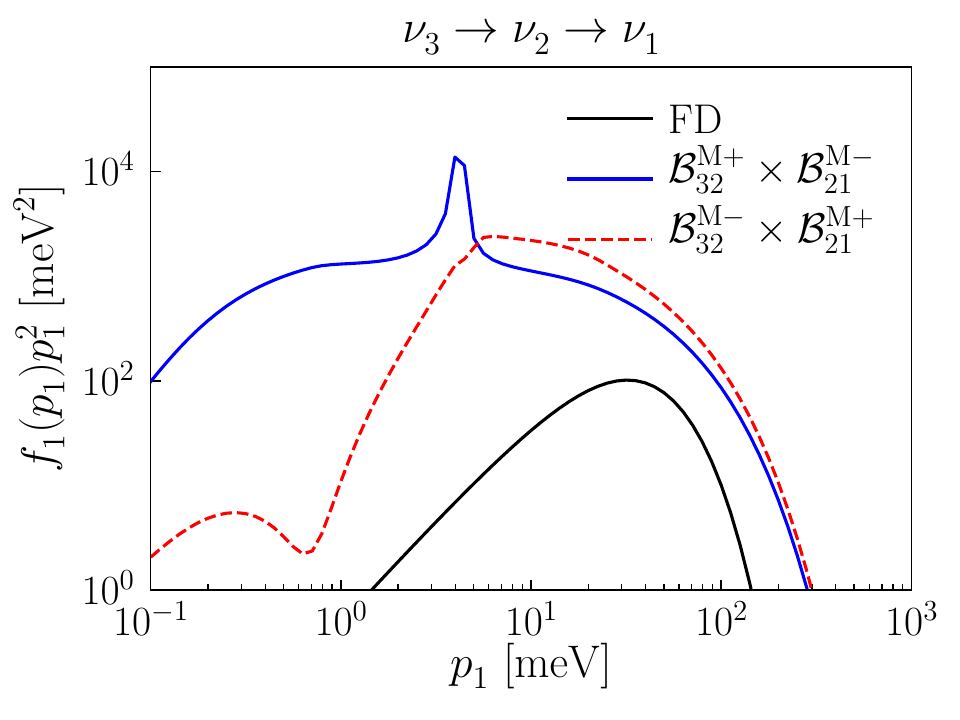}
	\vspace{-0.3cm}
	\caption{The distribution function $f^{}_1(p^{}_1) p^2_1$ of $\nu_1^{}$ from the decay $\nu_3^{} \to \nu_1^{} + \phi$ (upper left), $\nu_2^{} \to \nu_1^{} + \phi$ (upper right) and $\nu_3^{} \to \nu_2^{} \to \nu_1^{}$ (lower row). We assume the Majorana neutrino case with NO and choose $g=10^{-12}$ as the benchmark value. The FD distribution functions for primordial relic $\nu_1^{}$ when the decay happens are plotted in black curves. In the upper row, the distribution functions $f_1^{}(p_1^{}) p^2_1$ from helicity-preserving (${\cal B}_{ij}^{\rm M+}$) and -changing (${\cal B}_{ij}^{\rm M-}$) decays are plotted in blue solid curves and red dashed curves, respectively. In the lower row, we illustrate the energy spectra of daughter $\nu_1^{}$ from four different decay channels, including two helicity-preserving ones ($h_3^{} = h_1^{}$, left) and two helicity-changing ones ($h_3^{} = - h_1^{}$, right).}
	\label{fig:distribution}
\end{figure}

Considering the helicity-changing decay of Majorana neutrinos $\nu_i^{}(p^{}_i, \pm) \to \nu_j^{}(p^{}_j, \mp) + \phi$, we can easily find the distribution function of the final-state $\nu_j^{}$~\cite{Lipari:2007su}
\begin{eqnarray}
	\label{eq:distribution_produced_j}
	f_j^{}(E_j^{}) = \int_{l(E_j^{})}^{\infty} {\rm d}E_i^{} \ f_{i}^{} (E_i^{}) \times  \frac{1}{\Gamma_{\pm,i}^{\rm M}} \frac{{\rm d}\Gamma_{\pm \mp,ij}^{\rm M}}{{\rm d}E_j^{}} \;. 
\end{eqnarray}  
Here $f_{i}^{} (E_i^{})$ is the energy distribution function of parent neutrinos, and it is just the FD distribution function in Eq.~(\ref{eq:FD}) for the primordial $\nu^{}_i$. The lower limit of the integral in Eq.~(\ref{eq:distribution_produced_j}) is determined from $E_j^{\rm min}$ in Eq.~(\ref{eq:Ejmin}), i.e.,
\begin{eqnarray}
	l(E_j^{}) = \frac{E_j^{}}{2} \left(1+\frac{1}{r_{ji}^2}\right) + \frac{|{\bf p}_j^{}|}{2}\left(1-\frac{1}{r_{ji}^2}\right) \;.
\end{eqnarray}
For secondary decays, such as $\nu_3^{} \to \nu_2^{} \to \nu_1^{}$ in the NO case, the energy distribution function of $\nu_1^{}$ can be calculated similarly, after obtaining the energy spectrum of produced $\nu_2^{}$. To be more explicit, we plot the distribution function of Majorana neutrinos $\nu_1^{}$ in the NO case from the decay $\nu_3^{} \to \nu_1^{} + \phi$ (upper left), $\nu_2^{} \to \nu_1^{} + \phi$ (upper right) and $\nu_3^{} \to \nu_2^{} \to \nu_1^{}$ (lower row) in Fig.~\ref{fig:distribution}. The benchmark value of the coupling constant is $g = 10^{-12}$ and the corresponding $z_{3,2}^{\rm decay}$ can be directly read off from the upper left panel of Fig.~\ref{fig:g_zdecay}. We compare the spectra of produced $\nu_1^{}$ with the primordial FD distribution (black solid curves) at the same redshift. For the former two channels, the energy spectra from both helicity-preserving (blue solid curves) and -changing (red dashed curves) decays are plotted in order to illustrate their differences. On the other hand, there are four different channels to produce $\nu_1^{}$ with various helicities from the secondary decay of $\nu_2^{}$, which can be divided into the case of $h_3^{} = h_1^{}$ (lower left) and $h_3^{} = - h_1^{}$ (lower right). It can be observed that the energy spectrum can be regarded as a convolution of the FD distribution of primordial $\nu_3^{}$, the decay spectrum of $\nu_2^{}$ from $\nu_3^{} \to \nu_2^{} + \phi$ and that of $\nu_1^{}$ from $\nu_2^{} \to \nu_1^{} + \phi$.

As shown in Fig.~\ref{fig:A}, the velocities of parent neutrinos play a crucial role in the difference between helicity-changing and -preserving decays. To fully take such effects into account, any physical quantities depending on the neutrino velocity or three-momentum should be modified accordingly by considering their own energy spectrum. For instance, the neutrino number densities $n_i^\pm (z)$ and $n_j^\pm (z)$ in Eq.~(\ref{eq:daughter_number}) should be calculated as
\begin{eqnarray} 
	\label{eq:daughter_number_modified}
	n_i^{\pm} (z) &=& \frac{\displaystyle \int_{0}^{\infty} n_0^{} {\rm e}^{-\lambda_i^{} (z)}_{} \; f_{i}^{} \left[|{\bf p}^{}_i|,T_\nu^{}(z)\right] |{\bf p}^{}_i|^2\ {\rm d}|{\bf p}^{}_i|}{\displaystyle \int_{0}^{\infty} f_{i}^{} \left[|{\bf p}^{}_i|,T_\nu^{}(z)\right] |{\bf p}^{}_i|^2\ {\rm d}|{\bf p}^{}_i|} \;, \nonumber \\
	n_j^{\pm}(z) &=& \frac{\displaystyle \int_{0}^{\infty} n_0^{} \left[1 - {\rm e}^{-\lambda_i^{} (z)}_{} \right] {\cal B}^{\rm M+}_{ij} \; f_{j}^{} \left[|{\bf p}^{}_j|,T_\nu^{}(z)\right] |{\bf p}^{}_j|^2\ {\rm d}|{\bf p}^{}_j|}{\displaystyle \int_{0}^{\infty} f_{j}^{} \left[|{\bf p}^{}_j|,T_\nu^{}(z)\right] |{\bf p}^{}_j|^2\ {\rm d}|{\bf p}^{}_j|} \;.
\end{eqnarray}
Here $f_{i,j}^{}$ refers to the energy spectrum of $\nu_{i,j}^{}$, respectively. In the following discussions, we shall not explicitly indicate the integration over the distribution function, but just focus on physical quantities, which are in fact calculated numerically by considering the momentum distribution.

For illustration, we examine the decays of the Majorana neutrinos with the helicity $h = -1$ in the NO case. But our discussions can be easily extended to Dirac neutrinos or to the IO case. For a given three-momentum of the primordial neutrino $\nu^{}_i$ with $h=-1$ (denoted as $\nu^-_i$), the explicit expressions of neutrino number densities subject to invisible decays are as follows
\begin{eqnarray}
	n_3^-(z) &=& n_0^{} {\rm e}^{-\lambda_3^{} (z)}\;, \label{eq:n3-} \\
	n_2^-(z) &=& n_0^{} {\rm e}^{-\lambda_2^{} (z)} + \left[n_0^{} - n_3^-(z)\right] {\rm e}^{-\lambda_2^{\prime}(z)} {\cal B}^{\rm M+}_{32} \;, \label{eq:n2-}\\
	n_2^+(z) &=& \left[n_0^{} - n_3^-(z)\right] {\rm e}^{-\lambda_2^{\prime}(z)} {\cal B}_{32}^{\rm M-}  \;, \label{eq:n2+}\\
	n_1^-(z) &=& n_0^{} + \left[n_0^{} - n_3^-(z)\right] {\cal B}_{31}^{\rm M+}  + n_0^{} \left[1-{\rm e}^{-\lambda_2^{}(z)}\right] {\cal B}_{21}^{\rm M+} \nonumber \\
	&& + \left[n_0^{} - n_3^- (z)\right] {\cal B}_{32}^{\rm M+} \times \left[1-{\rm e}^{-\lambda_2^{\prime}(z)}\right] {\cal B}_{21}^{\rm M+} \nonumber \\
	&& + \left[n_0^{} - n_3^- (z)\right] {\cal B}_{32}^{\rm M-} \times \left[1-{\rm e}^{-\lambda_2^{\prime}(z)}\right]{\cal B}_{21}^{\rm M-} \;, \label{eq:n1-}\\
	n_1^+(z) &=& \left[n_0^{} - n_3^-(z)\right] {\cal B}_{31}^{\rm M-} + n_0^{} \left[1-{\rm e}^{-\lambda_2^{}(z)}\right] {\cal B}_{21}^{\rm M-} \nonumber \\
	&& + \left[n_0^{} - n_3^- (z)\right] {\cal B}_{32}^{\rm M+} \times \left[1-{\rm e}^{-\lambda_2^{\prime}(z)}\right] {\cal B}_{21}^{\rm M-} \nonumber \\
	&&  + \left[n_0^{} - n_3^- (z)\right] {\cal B}_{32}^{\rm M-} \times \left[1-{\rm e}^{-\lambda_2^{\prime}(z)}\right] {\cal B}_{21}^{\rm M+} \;. \label{eq:n1+}
\end{eqnarray}
The primordial neutrino number density $n_0^{} \approx 56~{\rm cm}^{-3}$ at the neutrino decoupling is set as usual. Some comments on the number densities of two lighter neutrinos are in order.
\begin{itemize}
	\item For $n_2^-(z)$ in Eq.~(\ref{eq:n2-}), the first term on the right-hand side represents the number density of primordial $\nu_2^{-}$ surviving the decays $\nu_2^{-} \to \nu_1^{\pm} +\phi$, similar to that for $\nu^-_3$ in Eq.~(\ref{eq:n3-}). The number density of newly produced $\nu_2^{-}$ from the helicity-preserving decay $\nu_3^{-} \to \nu_2^{-} + \phi$ is described by the second term with the branching ratio ${\cal B}_{32}^{\rm M+}$. These produced $\nu_2^-$ will undergo the secondary decays into $\nu_1^{}$ of both helicities, so the net number density should be multiplied by a factor of ${\rm e}^{-\lambda_2^{\prime}(z)}$. As the initial number density of $\nu_2^{+}$ is zero, they all come from the helicity-changing decays of $\nu_3^{-}$ with the branching ratio ${\cal B}_{32}^{\rm -}$.
	
	\item For $n_1^{-}(z)$ in Eq.~(\ref{eq:n1-}), the first three terms on the right-hand side account for the number density of the initial $\nu_1^{-}$, those produced from the helicity-preserving decays of $\nu_3^{-}$ and $\nu_2^{-}$, respectively. In addition, $\nu_1^{-}$ can also be generated from the secondary decays of $\nu_2^{}$ in two channels, namely, $\nu_3^{} \xrightarrow{{\cal B}_{32}^{\rm M+}} \nu_2^{} \xrightarrow{{\cal B}_{21}^{\rm M+}} \nu_1^{}$ and $\nu_3^{} \xrightarrow{{\cal B}_{32}^{\rm M-}} \nu_2^{} \xrightarrow{{\cal B}_{32}^{\rm M-}} \nu_1^{} $. The contributions from these two processes are given in the last two lines of Eq.~(\ref{eq:n1-}). The result for $n_1^{+}(z)$ in Eq.~(\ref{eq:n1+}) can be understood in a similar way.
\end{itemize}

Summing up all the number densities in Eqs.~(\ref{eq:n3-})-(\ref{eq:n1+}), one finds that the total number density of neutrinos is still $3n_0^{}$. This is what we expect since the total number of neutrinos is conserved during the decay processes. For the decays of Majorana neutrinos with the opposite helicity, all the calculations can be performed in the same way. Therefore, the total number densities of left- and right-helical neutrinos for each mass eigenstate are equal. 

\begin{figure}[t!]
	\centering
	\includegraphics[scale=0.54]{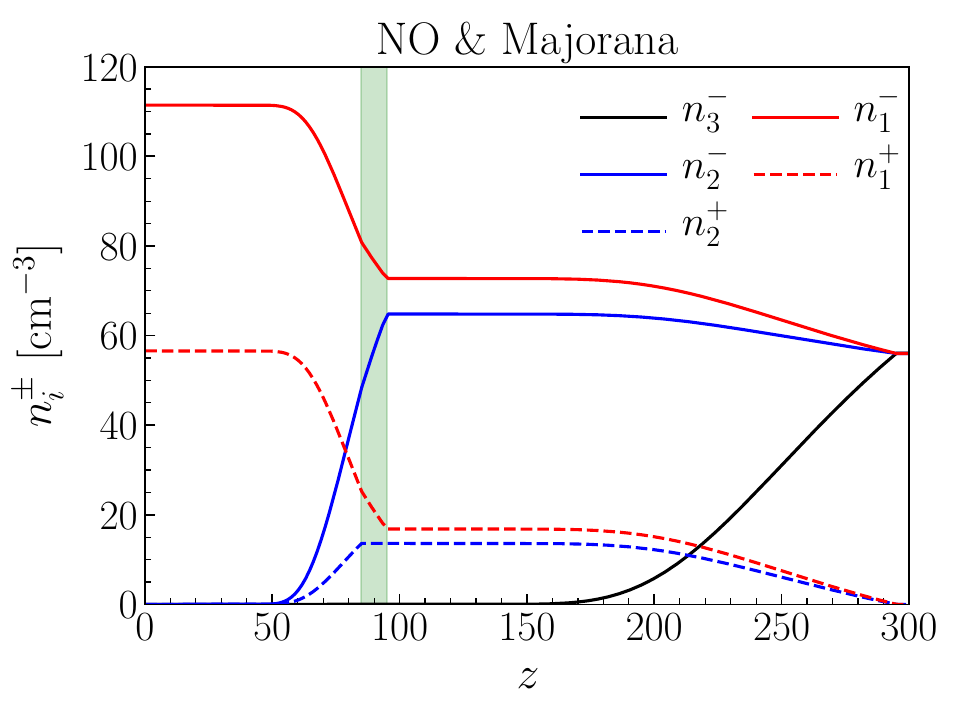} \hspace{-0.5cm}
	\includegraphics[scale=0.54]{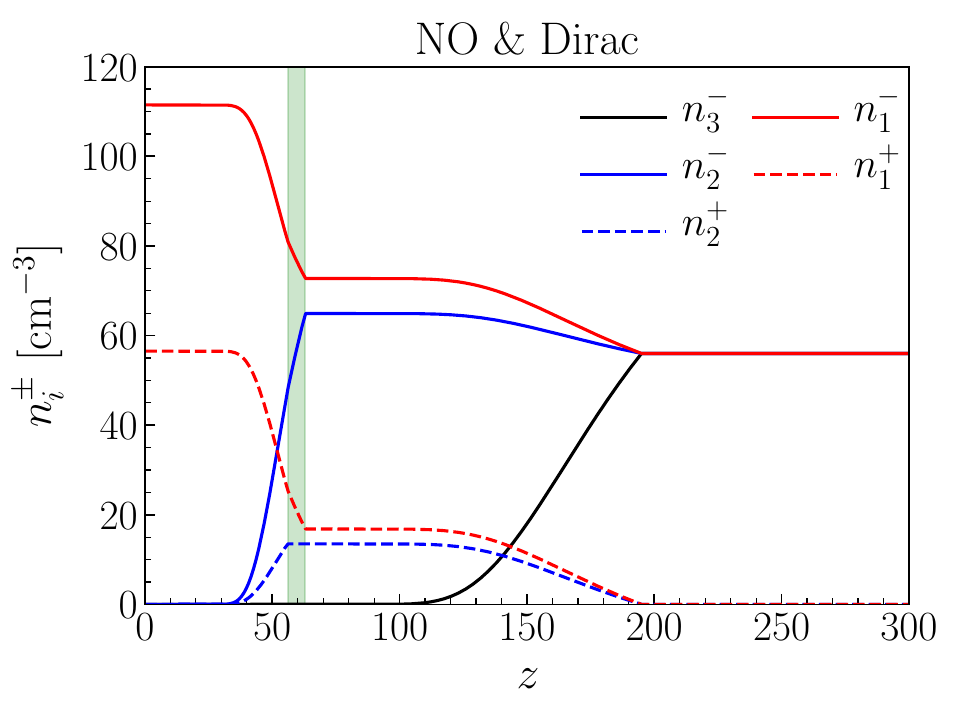} \\ \vspace{0.0cm}
	\includegraphics[scale=0.54]{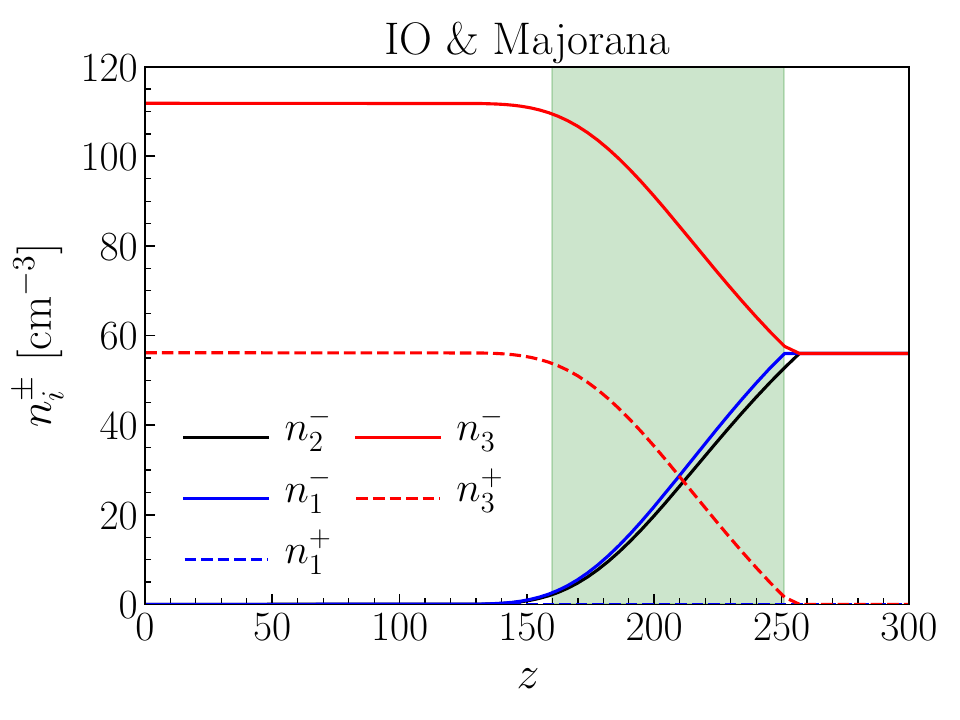} \hspace{-0.5cm}
	\includegraphics[scale=0.54]{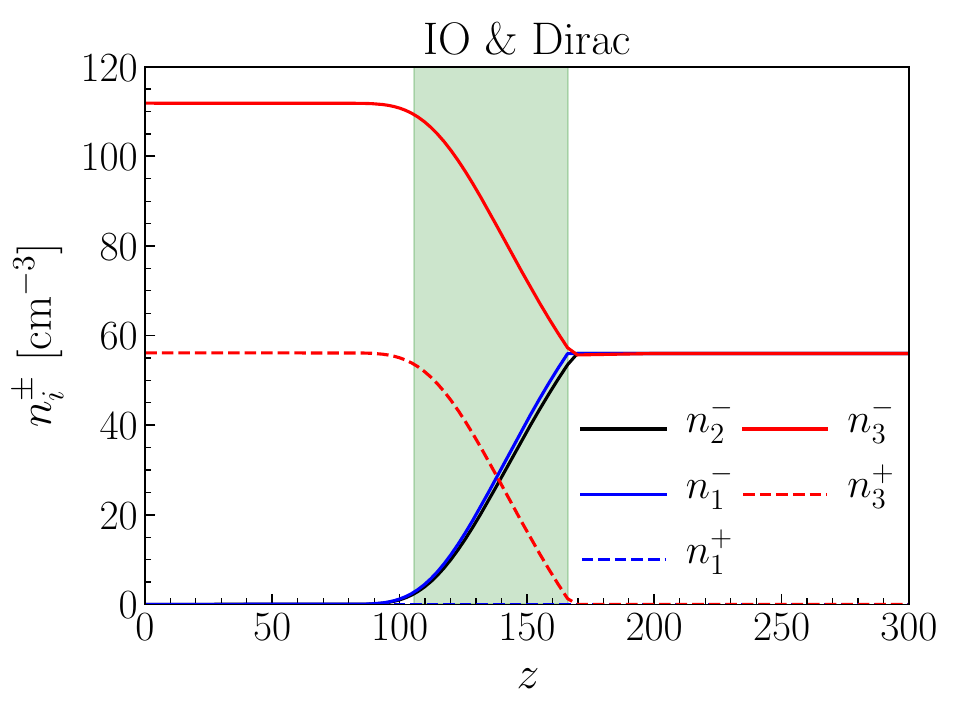}
	\vspace{-0.5cm}
	\caption{The evolution of neutrino number densities after invisible decays in the case of NO (upper row) and IO (lower row). To show the effects of the helicity-changing decays more intuitively, only decays from left-helical neutrinos are taken into consideration in the Majorana (left column) and Dirac case (right column). The benchmark value for the coupling constant is $g=10^{-12}$. Number densities of the heaviest ($\nu_3^{}$ in NO or $\nu_2^{}$ in IO), second-heaviest ($\nu_2^{}$ in NO or $\nu_1^{}$ in IO) and the lightest ($\nu_1^{}$ in NO or $\nu_3^{}$ in IO) neutrinos are plotted in black, blue and red curves, respectively, where left-helical ones are plotted in solid curves and right-helical ones are displayed in dashed curves. The green shadow highlights the time between the primary decay of primordial second-heaviest neutrinos and the secondary decay. The superscript ``$\pm$" represents the helicities $h=\pm 1$ of corresponding neutrinos.}
	\label{fig:z}
\end{figure}

The evolution of neutrino number densities with different helicities from the very beginning of decays till today is plotted in Fig.~\ref{fig:z}. The benchmark value of the coupling constant is chosen to be $g=10^{-12}$. As mentioned before, the number densities of left- and right-helical Majorana neutrinos are equal. To highlight the impact of helicity-changing decays, we only plot the evolution of primordial left-helical neutrinos and the produced ones from their decays (left column). Similarly, we also plot the evolution of left-helical neutrinos in the Dirac case (right column). The behavior of right-helical neutrinos in the Majorana case or the antineutrinos in the Dirac case can be described in the same way. The green shaded region denotes the period between the decay of the primordial second-heaviest neutrino and the subsequent decay of the produced ones. Some discussions on the evolutionary behavior in the Majorana case shown in Fig.~\ref{fig:z} are helpful.
\begin{itemize}
	\item In the NO case (left panel in the upper row), as the lifetimes of two unstable neutrinos (i.e., $\nu^{}_3$ and $\nu^{}_2$) differ significantly, there exists a production window for $\nu_2^{}$ (blue curves) during the decay of $\nu_3^{-}$ (black curve), creating both left-helical (solid curve) $\nu_2^{-}$ and right-helical $\nu_2^{+}$ (dashed curve). When the redshift reaches $z = z_2^{\rm decay} \approx 95$, corresponding to the lifetime of $\nu_2^{}$, the primordial $\nu_2^{-}$ starts to decay, leading to the decrease of $n_2^-$ and the further increase of $n_1^\pm$ (red curves). Later on, the secondary decay happens when $z \approx 85$, which can be directly seen from the falling of $n_2^{\pm}$ to almost zero and the increasing of $n_1^\pm$.
	
	\item In the IO case (left panel in the lower row), as two heavy neutrinos are almost degenerate in mass, the evolution behaviors (black and blue solid curves) are similar. In particular, the decay $\nu_2^{} \to \nu_1^{} + \phi$ is highly suppressed as its total rate $\Gamma_{\pm,21}^{\rm M} \propto (1-r_{12}^{})^3 \approx 0$ given $r_{12}^{} \approx 0.99$. Therefore, very few right-helical $\nu_1^{+}$ could be produced, as indicated by the blue dashed curve for $n_1^+$ always staying close to zero. On the other hand, as the lifetime of $\nu_2^{}$ is now longer than that of $\nu_3^{}$ in the NO case, the green shaded region becomes much wider than that in the NO case. 
\end{itemize}
It should be emphasized that the distribution function of the secondary neutrinos has been taken into account in our numerical calculations. Its impact on the number densities of neutrinos with different helicities can be remarkable.

\subsection{Capture Rates of C$\nu$B}

After calculating the number densities of different neutrino mass eigenstates with specific helicities in the present Universe, one can immediately derive the capture rates of C$\nu$B on the tritium target via $\nu_e^{} + {}^3{\rm H} \to e^{-} + {}^{3}{\rm He}$~\cite{Long:2014zva}
\begin{eqnarray} \label{eq:capture_rate}
	\Gamma_{\rm C\nu B}^{} = N_{\rm T}^{} \overline{\sigma} \sum_{i=1}^3 \sum_{h_i^{}=\pm 1} \left|U_{ei}^{}\right|^2 n_i^{}(h_i^{}) {\cal A}(\beta_i^{},h_i^{}) \;,
\end{eqnarray}
where $N_{\rm T}^{} $ is the number of tritium nuclei in the target. For the PTOLEMY-like experiment with a target mass of $m_{\rm T}^{} = 100~{\rm g}$~\cite{Betts:2013uya}, we have $N_{\rm T}^{} = m_{\rm T}^{} / m_{{}^3{\rm H}}^{} \approx 2\times 10^{25} $ with the tritium nuclear mass $m_{{}^3{\rm H}}^{} \approx 2808.92~{\rm MeV}$~\cite{Wang:2021xhn}. In addition, the cross section of the $\nu_i^{}$ capture is $\left|U_{ei}^{}\right|^2 \overline{\sigma}$, where $\overline{\sigma} \approx 3.834 \times 10^{-45}~{\rm cm}^2$ and $U$ is the Pontecorvo-Maki-Nakagawa-Sakata (PMNS) matrix~\cite{Pontecorvo:1957cp,Maki:1962mu} with the relevant matrix elements~\cite{Gonzalez-Garcia:2021dve}
\begin{eqnarray}
	\left|U_{e1}^{}\right|^2 \approx 0.677\;, \quad \left|U_{e2}^{}\right|^2 \approx 0.298\;, \quad \left|U_{e3}^{}\right|^2 \approx 0.023\;.
\end{eqnarray}
The helicity-dependent function reads ${\cal A}(\beta_i^{},h_i^{}) \equiv 1-h_i^{} \beta_i^{} $. In the non-relativistic limit $\beta_i^{} \to 0$, the function ${\cal A}(\beta_i^{},h_i^{}) \to 1$ is equal for both the left- and right-helical states. As a consequence, there is no need to distinguish between two helical states in the non-relativistic limit. However, for relativistic neutrinos when $\beta^{}_i \to 1$, different values of the helicity $h^{}_i$ will significantly change the value of ${\cal A}(\beta^{}_i, h^{}_i)$, which together with the helicity-dependent number density $n^{}_i(h^{}_i)$ ultimately affects the capture rates. As the lightest neutrino mass is chosen to be $m^{}_1 (m^{}_3) = 0.1~{\rm meV}$ in the NO (IO) case, we can safely neglect the gravitational clustering effects on massive neutrinos in the local galaxy. More related discussions can be found in Refs.~\cite{Ringwald:2004np,Zhang:2017ljh,Mertsch:2019qjv,Zimmer:2023jbb}. 

First, we briefly recall the difference in the capture rates for Dirac and Majorana neutrinos in the standard scenario of stable neutrinos.
\begin{itemize}
	\item In the Dirac case, only left-handed neutrinos and right-handed antineutrinos were created in the early Universe, where the helicity and chirality coincide in the relativistic limit. After the neutrino decoupling, neutrinos gradually become non-relativistic but their helicities are still unchanged. However, only left-helical neutrinos could be captured on the tritium. With $n_i^\pm = n_0^{} = 56~{\rm cm}^{-3}$ the capture rate of C$\nu$B in the PTOLEMY experiment is $\Gamma_{\rm C\nu B}^{\rm D} = n_0^{} N_{\rm T}^{} \overline{\sigma} \approx 4~{\rm yr}^{-1}$~\cite{Long:2014zva}. If we further take the momentum distribution into account, the capture rate becomes $\Gamma_{\rm C\nu B}^{\rm D} = n_0^{} N_{\rm T}^{} \overline{\sigma} \left(1+\sum_{i=1}^3 \left|U_{ei}^{}\right|^2 \left<\beta_i^{}\right>\right)$, where the average velocity of each neutrino mass eigenstate is given by
	\begin{eqnarray}\label{eq:average_velocity}
		\left<\beta_i^{}\right> = \frac{\displaystyle \int_0^\infty \beta_i^{}\ f_{\rm FD}^{}\left(\left|{\bf p}_i^{}\right|,T_{\nu_i^{}}^{0}\right)|{\bf p}_i^{}|^2\ {\rm d}|{\bf p}_i^{}|}{\displaystyle \int_0^\infty f_{\rm FD}^{}\left(|{\bf p}_i^{}|,T_{\nu_i^{}}^{0}\right)|{\bf p}_i^{}|^2\ {\rm d}|{\bf p}_i^{}|}\;.
	\end{eqnarray}
	Since the lightest neutrino $\nu^{}_1$ in the NO case can be ultra-relativistic today, it has been observed in Ref.~\cite{Roulet:2018fyh} that the capture rate can increase to $\Gamma_{\rm C\nu B}^{\rm D} \approx 7~{\rm yr}^{-1}$. In the IO case, as the PMNS matrix element associated with the lightest neutrino $\nu^{}_3$ is $|U_{e3}^{}|^2$, the capture rate will not be noticeably affected.
	
	\item In the Majorana case, both left- and right-helical neutrinos were produced in the early Universe, and nowadays both of them can interact with the target. Therefore, the capture rate $\Gamma_{\rm C\nu B}^{\rm M} \approx 8~{\rm yr}^{-1}$ is twice larger than that in the Dirac case~\cite{Long:2014zva}. Meanwhile, the summation over helicities on ${\cal A}(\beta_i^{},h_i^{})$ is independent of the velocity, i.e., ${\cal A}(\beta_i^{},+1) + {\cal A}(\beta_i^{},-1) =2 $, so the momentum distribution will not affect the total capture rate.
\end{itemize}

\begin{figure}[t]
	\centering
	\includegraphics[scale=0.54]{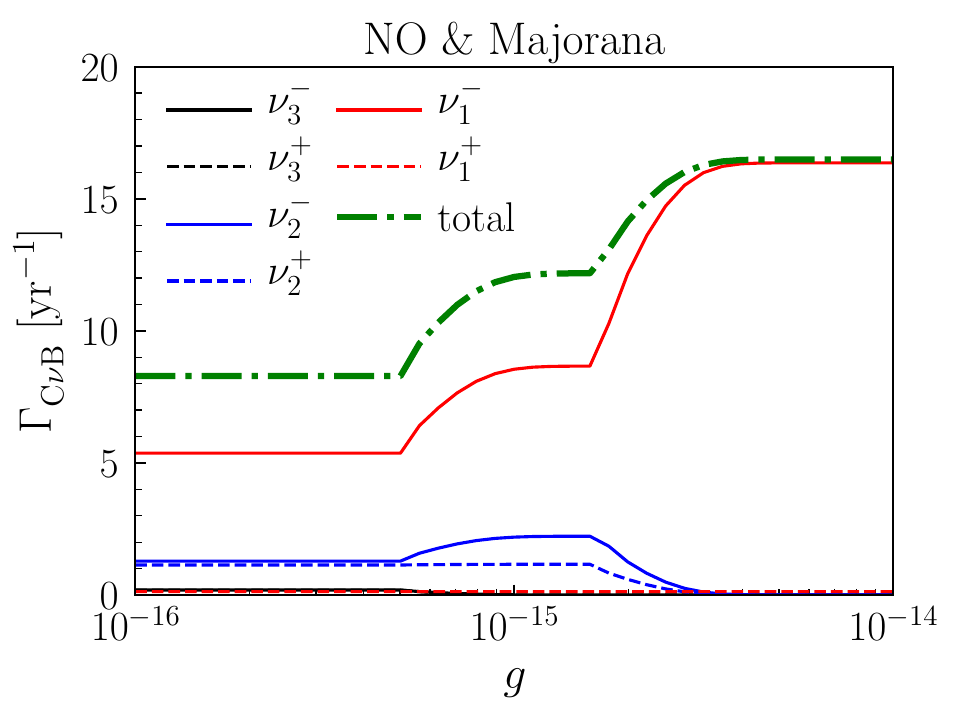} \hspace{-0.5cm}
	\includegraphics[scale=0.54]{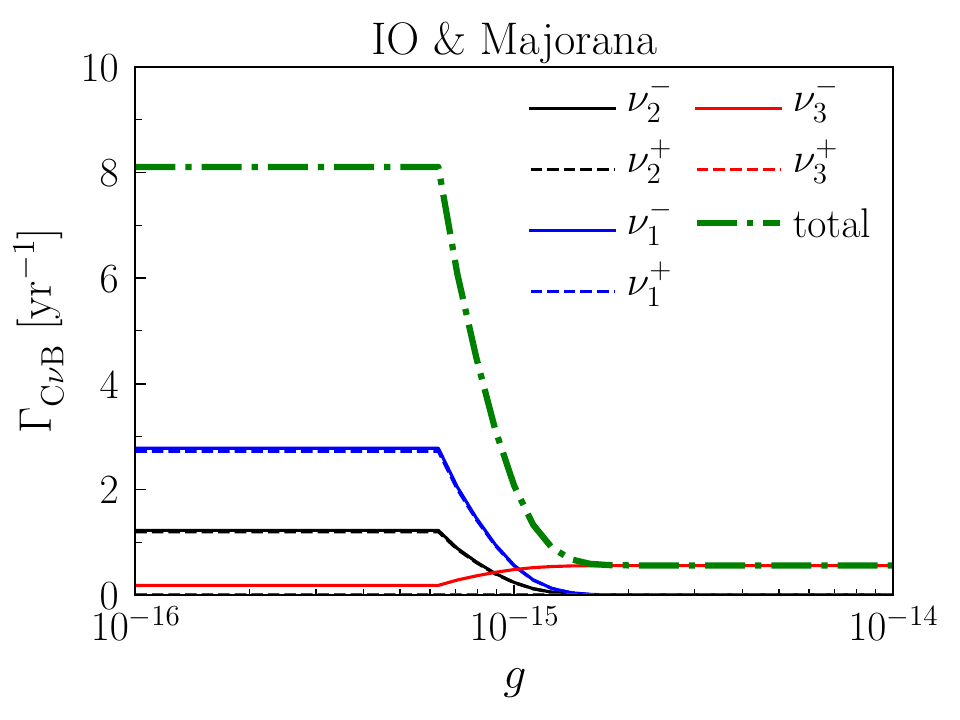} 
	\vspace{-0.5cm}
	\caption{The capture rate for Majorana neutrinos in the case of NO (left) and IO (right). The green dot-dashed curves represent the total capture rates in each case, while the contributions from each mass eigenstate with a specific helicity are plotted separately. Their legends are the same as those in Fig.~\ref{fig:z}.}
	\label{fig:cap_majorana}
\end{figure}

Then, we consider invisible decays of the C$\nu$B. The number densities can be calculated with Eqs.~(\ref{eq:n3-})-(\ref{eq:n1+}) by setting the redshift $z=0$, and consequently the capture rates will be different from those in the standard scenario. It is worthwhile to emphasize that the FD distribution $f_{\rm FD}^{}$ in Eq.~(\ref{eq:average_velocity}) should now be replaced by the distribution function $f_i^{}$ in Eq.~(\ref{eq:distribution_produced_j}) when calculating the average velocities of daughter neutrinos from invisible decays. Furthermore, the latter should also be evolved from the redshift when decays occur to the present time. Inserting the number densities with specific helicities into Eq.~(\ref{eq:capture_rate}), one can obtain the total capture rate.

\begin{figure}[t]
	\centering
	\includegraphics[scale=0.54]{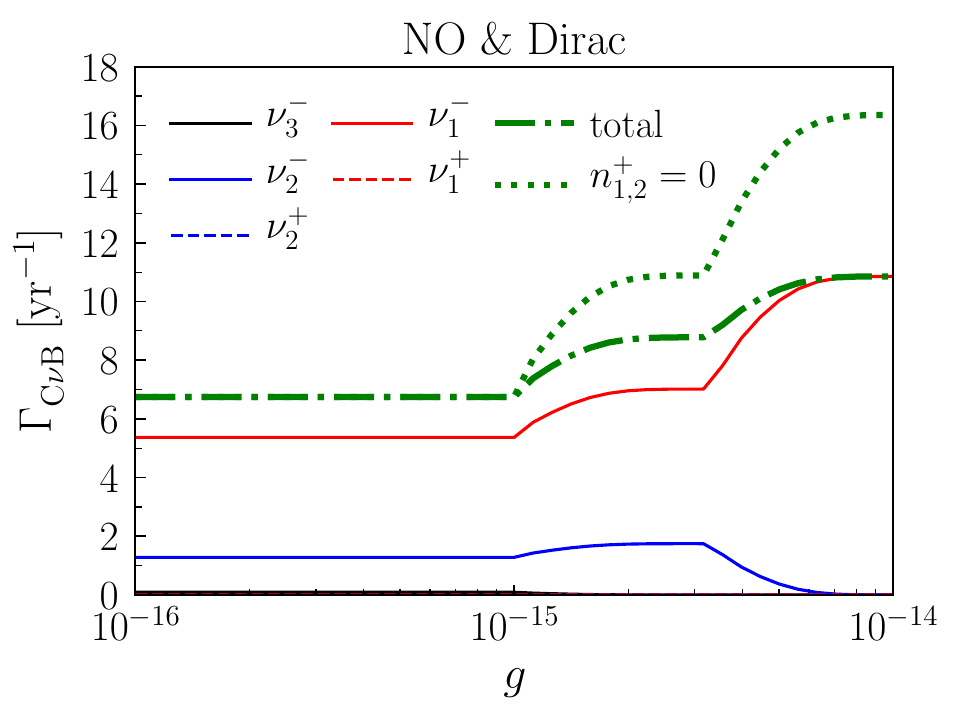} \hspace{-0.5cm}
	\includegraphics[scale=0.54]{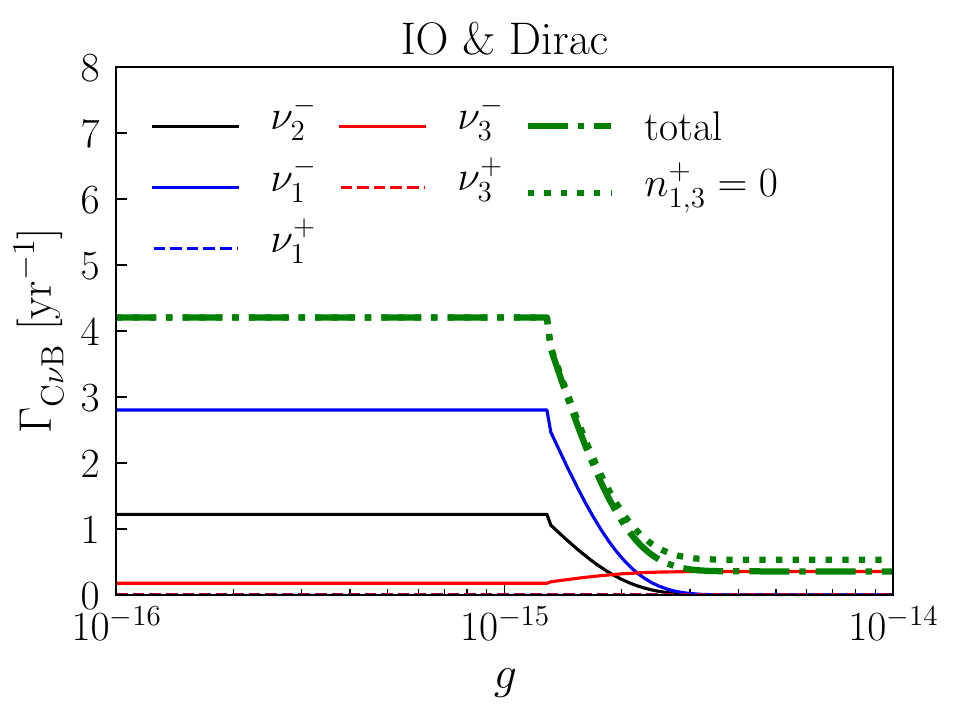}
	\vspace{-0.5cm}
	\caption{The capture rate for Dirac neutrinos in NO (left) and IO (right). To illustrate the effects of helicity-changing decays (green dot-dashed curves), we also plot the total capture rate in the green dotted curves, assuming all the neutrino decays to be helicity-preserving. Other legends are the same as those in Fig.~\ref{fig:z} and Fig.~\ref{fig:cap_majorana}.}
	\label{fig:cap_dirac}
\end{figure}

The numerical results of the total capture rate for various coupling constants $g$ are presented in Fig.~\ref{fig:cap_majorana} (for Majorana neutrinos) and Fig.~\ref{fig:cap_dirac} (for Dirac neutrinos) as green dot-dashed curves. It is clear from these figures that neutrino lifetimes are extremely long for very small couplings, so the results in the standard scenario will be reproduced. However, the main features of the capture rates for larger couplings deserve more detailed discussions.

\begin{itemize}	
	\item For Majorana neutrinos, the capture rate could reach $\Gamma_{\rm C\nu B}^{\rm M} \approx 16~{\rm yr}^{-1}$ in the case of NO for larger coupling constants, for which all heavier neutrinos have already decayed into the lightest one $\nu_1^{}$, while it returns to $\Gamma_{\rm C\nu B}^{\rm M} \approx 8~{\rm yr}^{-1}$ with smaller coupling constants as in the standard case. There are two conspicuous step-like increases. The first one comes from the contribution of $\nu_3^{}$ decays, accompanied by the increase of $n_2^{}$ and $n_1^{}$, and the second one is attributed to $\nu_2^{}$ decays. The contributions to the capture rate from $\nu_3^{}$ (black curve) are suppressed by the matrix element $|U_{e3}^{}|^2 \approx 0.023$, and those from $\nu_1^{+}$ (red dashed curve) are also suppressed by the function ${\cal A}(\beta_1^{},+1) = 1-\left<\beta_1^{}\right> \to 0$ with $\left<\beta_1^{}\right> \to 1$. Therefore, one can hardly observe those contributions from the plots. On the other hand, since the momentum distribution of $\nu_2^{}$ from decays no longer satisfies the FD distribution, the average velocity is not that small and causes certain differences on the capture rates for left- (blue solid curve) and right-helical (blue dashed curve) neutrinos as we mentioned before. In the IO case, as two heavy neutrinos are nearly degenerate in mass, there is no significant production of $\nu_1^{}$ from $\nu_2^{}$ decays. With larger coupling constants, since all heavier neutrinos have already decayed into $\nu_3^{}$, the capture rate is suppressed heavily by the matrix element $|U_{e3}^{}|^2$ and thus becomes as low as $\Gamma_{\rm C\nu B}^{\rm M} \approx 1~{\rm yr}^{-1}$.
	
	\item For Dirac neutrinos, the behavior of the capture rate as a function of the coupling constant is similar to that in the Majorana case. However, since only neutrinos can be captured, the total rates will be smaller than those for Majorana neutrinos. After considering the realistic distribution functions, we find that the total rate in the standard case is $\Gamma_{\rm C\nu B}^{\rm D} \approx 7~{\rm yr}^{-1}\ (4~{\rm yr}^{-1})$ in the NO (IO) case, which can be read off from Fig.~\ref{fig:cap_dirac} in the small-coupling limit. This is consistent with the results in Ref.~\cite{Roulet:2018fyh}. In the case of neutrino decays, the capture rate can reach $\Gamma_{\rm C\nu B}^{\rm D} \approx 11~{\rm yr}^{-1}$ for NO, while less than $1~{\rm yr}^{-1}$ for IO. 
	
\end{itemize}

In order to demonstrate the impact of helicity-changing neutrino decays, we also calculate the capture rate by assuming that neutrino decays are helicity-preserving. Under this assumption, there will be no change in the Majorana case, but the enhancement of the capture rate appears in the Dirac case. For comparison, the total capture rate without helicity-changing decays has been plotted in Fig.~\ref{fig:cap_dirac} as green dotted curves. In the large-coupling region (e.g., $g \sim 10^{-14}$), we have $\Gamma_{\rm C\nu B}^{\rm D} \approx 16~{\rm yr}^{-1}$ in the NO case (left panel). This observation can be understood by noticing that the number density of left-helical neutrino states will be much larger without helicity-changing decays. On the other hand, the difference becomes negligible in the IO case (right panel). The reason is simply that the decay rates of $\nu^{}_2 \to \nu^{}_1$ in this case are highly suppressed.

Finally, it is worth pointing out that the strategy adopted in this work may lead to some inaccuracy in the capture rate, compared to the approach of numerically solving the Boltzmann equations. In particular, when the coupling constant is around $g\sim {\cal O}(10^{-15})$, neutrinos may be still decaying nowadays. In this case, the inaccuracy arises from the fact that the characteristic redshift $z_i^{\rm decay}$ and the suppression factor ${\rm e}^{-\lambda_i^{}}$ are no longer valid to describe neutrino decays. The exact number densities of neutrinos can only be obtained by solving their evolution equations in the expanding Universe. A systematic study along this line will be left for future works.

\section{Summary}\label{sec:sum}

In this paper, we have carried out a thorough study of the invisible decays of massive neutrinos and investigated their implications for the experimental detection of the C$\nu$B neutrinos. First, the decay rates in the helicity-preserving and -changing decay channels are calculated and discussed in detail. The branching ratios of the helicity-changing decays can be significant, depending on the neutrino mass spectrum and the specific decay processes. Second, by taking the benchmark value $g = 10^{-12}$ of the coupling between massive neutrinos and the Nambu-Goldstone boson, we explain how to evaluate the neutrino number densities in an effective way. Furthermore, we have found that the distribution function of the secondary neutrinos from decays could remarkably deviate from the Fermi-Dirac one. Third, considering neutrino decays and the distribution function of the lightest neutrino, we obtain the capture rates of C$\nu$B in the PTOLEMY-like experiment. Quite different from the case of stable neutrinos, the capture rates in the PTOLEMY-like experiments increase to $\Gamma_{\rm C\nu B}^{\rm M} \approx 16~{\rm yr}^{-1}$ for Majorana neutrinos and $\Gamma_{\rm C\nu B}^{\rm D} \approx 11~{\rm yr}^{-1}$ for Dirac neutrinos in the NO case, whereas those in the IO case are about one event per year. Nevertheless, detecting C$\nu$B would be extremely difficult if all relic neutrinos have already decayed into the lightest mass eigenstate nowadays, no matter whether neutrinos are Dirac or Majorana particles in the NO or IO case. The main reason is that the energy resolution $\Delta \lesssim 0.7 m_\nu^{} $ is required for the PTOLEMY-like experiments~\cite{Long:2014zva}. With the lightest neutrino mass $m_{\rm lightest}^{}=0.1~{\rm meV}$ in our discussions, which should be compared with the expected energy resolution $\Delta \approx 50~{\rm meV}$ in future experiments~\cite{PTOLEMY:2018jst,PTOLEMY:2019hkd}, it will be practically impossible to distinguish the C$\nu$B signal from the tritium beta-decay spectrum.

Although the detection of C$\nu$B neutrinos is definitely one of the most challenging tasks in particle physics and cosmology, it is extremely important for us to probe the intrinsic properties of massive neutrinos, such as their Dirac or Majorana nature, absolute mass scale and lifetimes. In fact, the PTOLEMY-like experiments have been designed with the capability to measure the absolute neutrino mass and detect these relic neutrinos from the Big Bang. In this sense, we believe that our calculations in this work are not only beneficial to deeply understanding basic properties of massive neutrinos, but also serve as an instructive example for testing new-physics scenarios beyond the SM in future experiments.

\section*{Acknowledgments}

The authors would like to thank Prof. Xiao-Gang He for helpful discussions on neutrino decays. This work was supported by the National Natural Science Foundation of China under grant No.~11835013 and the CAS Project for Young Scientists in Basic Research (YSBR-099).

\bibliographystyle{elsarticle-num}
\bibliography{ref_CvB}

\end{document}